\documentclass[useAMS,usenatbib]{mn2e}

\usepackage{amssymb}
\usepackage{times}
\usepackage{graphicx}
\usepackage{epstopdf}
\usepackage{subfigure}
\usepackage[T1]{fontenc}
\usepackage{aecompl} 
\usepackage{multirow}
\usepackage{hyperref}

\newcommand{\beq}{\begin{equation}}
\newcommand{\eeq}{\end{equation}}

\def\gs{\mathrel{\lower0.6ex\hbox{$\buildrel {\textstyle >}\over{\scriptstyle \sim}$}}}
\def\ls{\mathrel{\lower0.6ex\hbox{$\buildrel {\textstyle <}\over{\scriptstyle \sim}$}}}
\newcommand{\simgt}{\lower.5ex\hbox{$\; \buildrel > \over \sim \;$}}
\newcommand{\simlt}{\lower.5ex\hbox{$\; \buildrel < \over \sim \;$}}

\newcommand{\aap}{A\&A}
\newcommand{\apj}{ApJ}

\newcommand{\apjs}{ApJS}
\newcommand{\aj}{AJ}

\newcommand{\mnras}{MNRAS}

\newcommand{\ssr}{Space Science Reviews}

\defcitealias{se+et14_comalit_I}{CoMaLit-I}
\defcitealias{ser+al14_comalit_II}{CoMaLit-II}
\defcitealias{ser14_comalit_III}{CoMaLit-III}
\defcitealias{se+et14_comalit_IV}{CoMaLit-IV}

\begin{document}

\title[CoMaLit-I]{Comparing Masses in Literature (CoMaLit)-I. Bias and scatter in weak lensing and X-ray mass estimates of clusters}
\author[
M. Sereno, S. Ettori
]{
Mauro Sereno$^{1,2}$\thanks{E-mail: mauro.sereno@unibo.it (MS)}, Stefano Ettori$^{2,3}$
\\
$^1$Dipartimento di Fisica e Astronomia, Alma Mater Studiorum -- Universit\`a di Bologna, viale Berti Pichat 6/2, 40127 Bologna, Italia\\
$^2$INAF, Osservatorio Astronomico di Bologna, via Ranzani 1, 40127 Bologna, Italia\\
$^3$INFN, Sezione di Bologna, viale Berti Pichat 6/2, 40127 Bologna, Italia\\
}


\maketitle

\begin{abstract}
The first building block to use galaxy clusters in astrophysics and cosmology is the accurate determination of their mass. Two of the most well regarded direct mass estimators are based on weak lensing (WL) determinations or X-ray analyses assuming hydrostatic equilibrium (HE). By comparing these two mass measurements in samples of rich clusters, we determined the intrinsic scatters, $\sigma_\mathrm{WL}\sim$15 per cent for WL masses and $\sigma_\mathrm{HE}\sim$25 per cent for HE masses. The certain assessment of the bias is hampered by differences as large as $\sim$40 per cent in either WL or HE mass estimates reported by different groups. If the intrinsic scatter in the mass estimate is not considered, the slope of any scaling relation `observable--mass' is biased towards shallower values, whereas the intrinsic scatter of the scaling is over-estimated. 
\end{abstract}


\section{Introduction}
\label{sec_intr}

Usage of clusters of galaxies in cosmology and astrophysics relies on precise determination of their masses \citep{voi05,lim+al13}. In the context of ongoing and future large surveys \citep{eucl_lau_11}, cluster properties which can be easily measured, e.g., optical richness, X-ray luminosity, Sunyaev-Zel'dovich (SZ) flux, are used as mass proxies. This requires an accurate calibration of the observable through comparison with direct mass estimates \citep{an+be12,ett13}. 

The assessment of scaling relations is the foundation for investigating the physics of the baryons and of the dark matter at the cluster scale \citep{pra+al09,arn+al10,gio+al13}. Cosmological parameters can be constrained with cluster abundances and the observed growth of massive galaxy clusters \citep{vik+al09b,man+al10,planck_2013_XX} or with gas fractions \citep{ett+al09}.

Two of the most well regarded mass estimates are the weak lensing (WL) mass and the X-ray mass exploiting hydrostatic equilibrium (HE). Weak lensing observations of the shear distortion of background galaxies trace the gravitational field of the matter distribution of the lens \citep{hoe+al12,wtg_I_14,ume+al14}. The physics behind gravitational lensing is very well understood and WL provides unbiased estimates of the total mass along the line os sight. The problem is to single out the contribution of the lens and to de-project the information to get the intrinsic mass, which can then be confronted with theoretical predictions.

Under the assumption that hydrostatic equilibrium holds between the intracluster medium (ICM) and the gravitational potential, the cluster mass can be recovered from observations of the spatially resolved spectroscopic data and the X-ray surface brightness \citep{lar+al06,don+al14}. However, deviations from equilibrium or non-thermal contributions to the pressure are difficult to quantify and can bias the mass estimate.

Methods based on spectroscopic measurements of galaxies velocities, such as the caustic technique \citep{ri+di06} or approaches exploiting the Jeans equation \citep{lem+al09,biv+al13}, can be effective too but they are hindered by the expensive observational requirements and are mostly limited to low redshift halos.

In principle either WL or HE can provide accurate and unbiased mass measurements, but the approximations that have to be used (e.g., spherical symmetry, smooth density distributions, thermal equilibrium) may bias and scatter the results. These effects must be accurately quantified to calibrate other mass proxies. Due to scatter, WL and HE masses are proxies to the true mass themselves.

Numerical studies argued that lensing masses obtained from the fit of the cluster tangential shear profiles with Navarro-Frenk-White \citep[ NFW] {nfw96} functionals are biased low by $\sim$5--10 per cent with a scatter of $\sim$10--25 per cent \citep{men+al10,be+kr11,ras+al12}. The main sources of uncertainty in deprojected WL mass measurements are due to the presence of substructures and triaxiality. Lensing properties depend on the orientation of the cluster with respect to the line of sight \citep{ogu+al05,ser07,se+um11,lim+al13}. For systems whose major axis points toward the observer, 3D masses derived under the standard assumption of spherical symmetry are typically over-estimated. The opposite occurs for clusters elongated in the plane of the sky, which are in the majority if the selected sample is randomly oriented. 

The presence of substructures in the cluster surroundings may dilute the tangential shear signal \citep{men+al10,gio+al12a,gio+al14}. Severe mass under-estimations may come from either massive sub-clumps \citep{men+al10} or uncorrelated large-scale matter projections along the line of sight \citep{be+kr11}.

The scatter is less significant in optimally selected clusters either having regular morphology or living in substructure-poor environments \citep{ras+al12}. 

The origins of bias and scatter of X-ray masses are well understood too, even though they are difficult to quantify \citep{ras+al12}. They are strictly connected to non-thermal sources of pressure in the gas, to temperature inhomogeneity, and, to a lesser degree and mainly in the external regions, to the presence of clumps. Even if the cluster is in hydrostatic equilibrium, the assumption that all the pressure is thermal biases the HE mass low. Large-scale, unvirialised bulk motions and subsonic turbulence contribute kinetic pressure \citep{bat+al12}. 

Furthermore, structures in the temperature distribution bias low the temperature estimate. In fact, the X-ray detectors of Chandra and XMM-Newton (X-ray Multi-Mirror Mission) have a higher efficiency in the soft band and, thus, weight more colder gas \citep{maz+al04}. 

Numerical simulations showed that X-ray masses based on hydrostatic equilibrium are biased low by a large amount of $\sim$25--35 per cent \citep{pi+va08,ras+al12,ras+al14}. The bias grows from the inner to the outer regions of the clusters, where the presence of non-thermal sources of pressure in the ICM and temperature inhomogeneity play a larger role \citep{ras+al12}. 

Since the intrinsic scatters in either WL or HE masses have different origins, they are mostly uncorrelated. Scatter in WL masses is mainly due to triaxiality and sub-structures in the dark matter halo. On the other hand, the gas distribution approximately follows the gravitational potential and it is rounder than the dark matter one. Dark matter substructures are not necessarily associated to gas clumps. The sources which cause scatter in the HE masses are more related to gas physics and temperature distributions than to the total matter distribution and have a small impact on WL estimates.

On the observational side, the certain assessment of cluster masses is further complicated by instrumental and methodological sources of errors which may cause systematic uncertainties in data analysis \citep{roz+al14}.

The main sources of systematics in WL masses are due to selection and redshift estimate of background galaxies, which can be obtained through accurate photometric redshifts and colour-colour selection methods \citep{med+al10}, and to the calibration of the shear signal. A small calibration correction of the order of just a few percents translates into a typical error of $\sim$ 10 per cent in the estimate of the virial mass \citep{ume+al14}.

Instrumental uncertainty has long been recognised as one of the main source of systematics plaguing HE masses. XMM cluster temperatures are systematically smaller by 10-20 per cent than Chandra estimates \citep{nev+al10,don+al14}. On the other hand, Chandra and XMM measurements of the gas distribution are highly consistent with one another \citep{roz+al14,don+al14}. 

The picture on the inconsistencies between Chandra and XMM results is still debated. \citet{don+al14} found that Chandra and XMM temperatures of the very massive CLASH  \citep[Cluster Lensing And Supernova survey with Hubble,][]{pos+al12} clusters agree in the core, where photon fluxes are considerable, whereas the regions where the temperature differences are larger are typically $\sim$1 arcmin from the much brighter cluster core. Temperature differences persist even in outer regions with large signal-to-background ratio. These temperature discrepancies caused analogue off-sets in the HE mass.

\citet{mar+al14} compared the mass profiles of 21 LoCuSS (Local Cluster Substructure Survey) clusters that were observed with both satellites, extracting surface brightness and temperature profiles from identical regions of the respective datasets and including analytic models that predict the spatial variation of the Chandra and XMM-Newton backgrounds to $\la$2 and  $\la$5 per cent precision, respectively. Notwithstanding global XMM spectroscopic temperatures lower by $\sim$ 10 per cent, they obtained consistent results for the gas and total hydrostatic cluster masses. \citet{mar+al14} explained this counterintuitive result noticing that temperature discrepancies were significant only above a value of 6 keV. In the outer regions, most of the estimated temperatures were lower than this threshold and Chandra and XMM temperatures were in good agreement. Furthermore, they argued that larger errors bars are associated to highest temperature, due to the larger difficulty to distinguish the hottest spectra having a flatter shape from the background. The relative statistical weight in a fitting procedure is then lower. 

This is the first in a series of papers focused on COmparing MAsses in LITerature (CoMaLit). Here, we look for systematic differences in WL and HE masses obtained from independent analyses and we assess the overall level of bias and intrinsic scatter. According to numerical simulations, the scatter in X-ray masses is supposedly smaller than in weak-lensing masses but a definite assessment of the values of bias and scatter of HE masses is still lacking, due to uncertainties in the treatment of the gas physics and to variability caused by the hydrodynamical scheme adopted in numerical simulations \citep{ras+al14}. In this paper, we provide the first measurements of intrinsic scatters of WL and HE masses of real clusters.

In the second paper of the series \citep[ CoMaLit-II]{ser+al14_comalit_II}, the Bayesian method developed to calibrate scaling relations between masses and observables, which fully accounts for intrinsic scatters in both the mass estimate and the scaling relation, was applied to the scaling relation between Sunyaev-Zel'dovich (SZ) flux and mass in Planck selected clusters of galaxies \citep{planck_2013_XXIX}. The third paper of the series \citep[ CoMaLit-III]{ser14_comalit_III} presents the Literature Catalogs of weak Lensing Clusters of galaxies (LC$^2$),  which are standardised compilations of clusters with measured WL masses. The fourth paper of the series \citep[ CoMaLit-IV]{se+et14_comalit_IV} extends the Bayesian methodology to account for time-evolution of the scaling relation. Products associated with the CoMaLit series, as well as future updates, will be hosted at \url{http://pico.bo.astro.it/\textasciitilde sereno/CoMaLit/}.

The present paper is structured as follows. In Sec.~\ref{sec_bias}, we discuss how the scatter in mass proxies can be estimated and how it impacts the calibration of scaling relations. Samples of clusters used in the analysis are introduced in Sec.~\ref{sec_samp}. Comparison among either WL or HE masses from different groups is investigated in Sec.~\ref{sec_mass_comp}. Section~\ref{sec_resu} is devoted to the measurements of scatter and bias affecting the mass proxies. Discussion of results is contained in Sec.~\ref{sec_disc}. Final considerations can be found in Sec.~\ref{sec_conc}. Three appendices contain some simplified and ready-to-use formulae to de-bias the scaling relations. A toy-model to illustrate the bias induced by the intrinsic scatter in the mass estimate is discussed in Appendix~\ref{app_bias}. The correction for the widely used Bivariate Correlated Errors and Intrinsic Scatter method \citep[BCES,][]{ak+be96} is proposed in Appendix~\ref{app_bces}. Appendix~\ref{app_mass} details how WL and HE mass estimates depends on the cosmological parameters.

Throughout the series of papers, we assume a fiducial flat $\Lambda$CDM cosmology with density parameter $\Omega_\mathrm{M}=0.3$, and Hubble constant $H_0=70~\mathrm{km~s}^{-1}\mathrm{Mpc}^{-1}$; $M_\Delta$ denotes the mass within the radius $r_\Delta$, which encloses a mean over-density of $\Delta$ times the critical density at the cluster redshift, $\rho_\mathrm{cr}=3H(z)^2/(8\pi G)$; $H(z)$ is the redshift dependent Hubble parameter. When $H_0$ is not specified, $h$ is the Hubble constant in units of $100~\mathrm{km~s}^{-1}\mathrm{Mpc}^{-1}$.

The presence of the superscript `WL', `HE', and `Tr', means that $M_{500}$ and $r_{500}$ were determined using the mass estimate from the WL analysis, the X-ray measurements, or the knowledge of the true mass (which is available only for simulated clusters), respectively. `$\log$' is the logarithm to base 10 and `$\ln$' is the natural logarithm.

\section{Biases and scatter induced biases}
\label{sec_bias}

\begin{figure*}
\begin{tabular}{cc}
\includegraphics[width=8.cm]{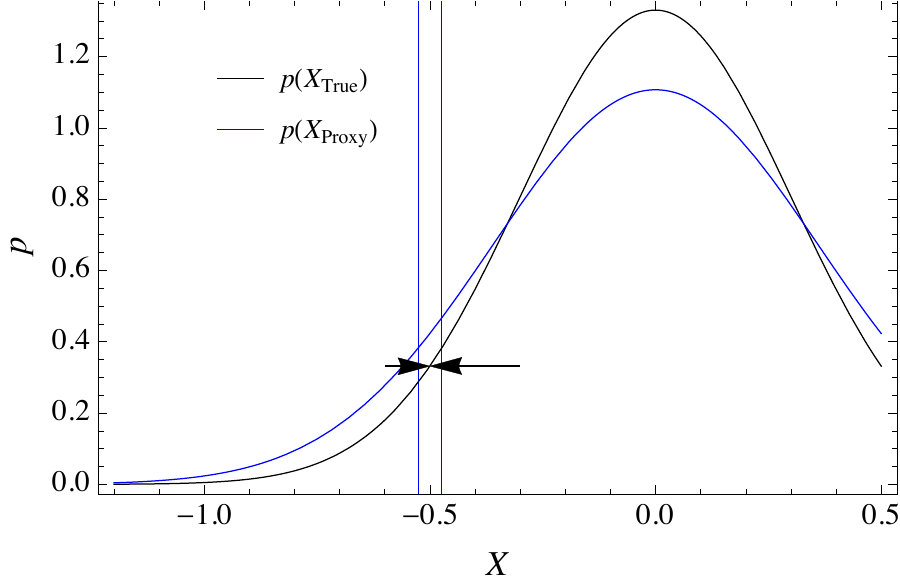} &
  \includegraphics[width=8.cm]{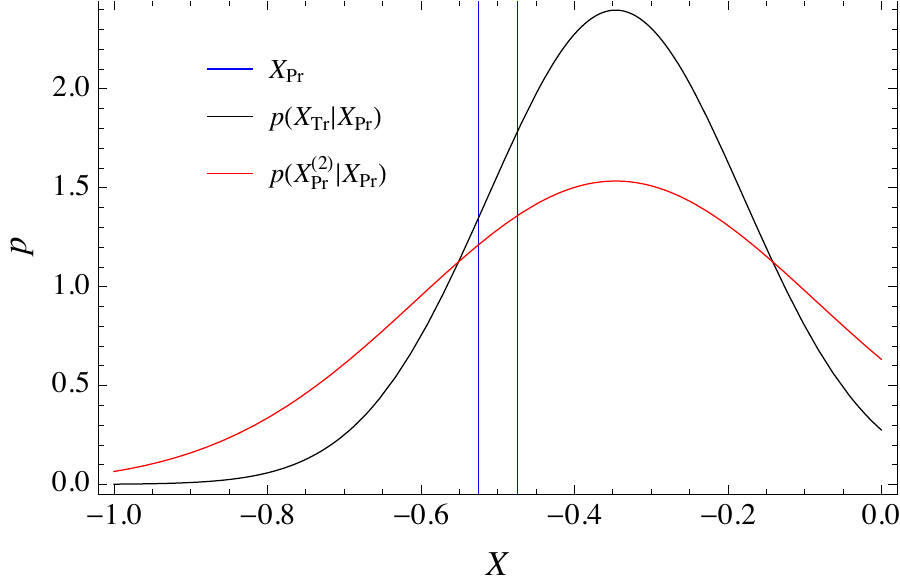} 
 \end{tabular}
\caption{Probability distributions of a quantity $X_\mathrm{True}$ and of its scattered proxy $X_\mathrm{Proxy}$. {\it Left panel}: the probability distribution $p(X_\mathrm{True})$ is a Gaussian function. $p(X_\mathrm{Proxy})$ is Gaussian too. It is smoothed due to the intrinsic (normal) scatter. At the tail at low values, for samples selected according to $X_\mathrm{Proxy}$ (interval delimited by the blue vertical lines), more objects with larger $X_\mathrm{True}$ are scattered into the subsample from the right side than from the left side. {\it Right panel}: conditional probability of $X_\mathrm{True}$ given $X_\mathrm{Proxy}$, $p(X_\mathrm{True}|X_\mathrm{Proxy})$ (black line), and of a second proxy $X_\mathrm{Proxy}^{(2)}$ given $X_\mathrm{Proxy}$, $p(X_\mathrm{Proxy}^{(2)}|X_\mathrm{Proxy})$ (red line). The scatters in $X_\mathrm{Proxy}$ and $X_\mathrm{Proxy}^{(2)}$ are not correlated. The mean $X_\mathrm{True}$ given $X_\mathrm{Proxy}$ is larger than $X_\mathrm{Proxy}$ ,  $\langle X_\mathrm{True} \rangle (X_\mathrm{Proxy})>X_\mathrm{Proxy}$. The second proxy $X_\mathrm{Proxy}^{(2)}$ is unbiased with respect to $X_\mathrm{True}$.}
\label{fig_bias}
\end{figure*}

The biases and the scatters of two mass proxies can be estimated by comparing the proxies in a cluster sample. The lensing and the hydrostatic mass approximate the true mass as
\begin{eqnarray}
\ln M^\mathrm{WL} \pm  \delta_\mathrm{WL} & = & \alpha_\mathrm{WL} + \beta_\mathrm{WL} \ln M^\mathrm{Tr} \pm \sigma_\mathrm{WL} \label{eq_proxy_WL}, \\
\ln M^\mathrm{HE} \pm  \delta_\mathrm{HE}  & = &  \alpha_\mathrm{HE} + \beta_\mathrm{HE} \ln M^\mathrm{Tr} \pm \sigma_\mathrm{HE} \label{eq_proxy_HE},  
\end{eqnarray}
where the $\alpha$'s quantify the bias and the $\beta$'s embody any deviation from linearity. The intrinsic scatters $\sigma_\mathrm{WL}$ and $\sigma_\mathrm{HE}$ are due to different physical processes and are assumed to be uncorrelated, see Sec.~\ref{sec_intr}. The actual WL (HE) mass is known save for a measurement error $\delta_\mathrm{WL}$ ($\delta_\mathrm{HE}$).

The notation in the left sides of Eqs.~(\ref{eq_proxy_WL},~\ref{eq_proxy_HE}) is a shortcut to remind that the measurement processes are affected by errors and uncertainties. The observed masses differ from the `true' values of the WL and HE masses which we would measure in the ideal case of infinitely accurate and precise observations and  systematics free-analyses. The notation in the right sides of Eqs.~(\ref{eq_proxy_WL},~\ref{eq_proxy_HE}) reminds us that even in the case of ideal measurements without systematic/statistical errors ($\delta=0$), the  `true' WL or HE masses differ from the `true' mass of the cluster due to intrinsic scatter. In this sense, WL and HE masses are mass proxies. The relations in either the left or the right side of Eqs.~(\ref{eq_proxy_WL},~\ref{eq_proxy_HE}) can be modelled with normal distributions.

Bias and scatter in logarithmic variables differ from analogue quantities in linear variables. We adopt (natural) logarithmic variables for coherence with the standard derivation of scaling relations.

\subsection{Eddington-like bias}
\label{sec_bias_edd}

Intrinsic scatter in the mass proxy induces systematic effects alike to the Eddington bias \citep{edd13,jef38,edd40}, which was first discussed in relation to observational uncertainties.
Due to scatter and even in absence of measurement errors, the average value of an observed proxy differs from the true intrinsic mean for objects of the same class, see Fig.~\ref{fig_bias}. When a subsample is selected according to the measured values of the proxy, $X_\mathrm{Proxy}$, the distribution of the differences between the proxy and the true values, $X_\mathrm{True}$, may be biased. 

For quantities drawn from a limited range, border and selection effects have to be considered. Near a threshold, the symmetry between objects that are scattered into a range of observed values from above and objects that are scattered into from below is broken. This can be accounted for by assuming that the true masses are drawn from a normal rather than a uniform distribution. 

Let us assume that the proxy $X_\mathrm{True}$ is the WL mass. Due to selection effects, the observed sample may be poor in clusters below a given threshold. At the tail at low values, more objects with larger $X_\mathrm{True}$ are scattered into the subsample from the right side, than from the less populated left side where the $X_\mathrm{True}$'s are smaller, see Fig.~\ref{fig_bias}. In a sample with steep bounds in true mass, clusters with very low values of  $M^\mathrm{WL}$ are then of two main kinds. They are either intrinsically less massive clusters, i.e., with low values of $M^\mathrm{Tr}$ and nearly unbiased values of $M^\mathrm{WL}$, or clusters with higher values of $M^\mathrm{Tr}$ that are scattered to lower values of observed $M^\mathrm{WL}$. The mean $M^\mathrm{Tr}$ is larger than the measured $M^\mathrm{WL}$.

Let us consider a second proxy such as the HE mass, whose intrinsic scatter is uncorrelated to the WL mass. The second proxy is not biased by selections based on the first proxy. If the clusters are selected according to their $M^\mathrm{WL}$, $M^\mathrm{HE}$ is still an unbiased scattered proxy of the true mass, since the scatters in measurements of WL and X-ray masses are uncorrelated. As a consequence, the mass ratio $M^\mathrm{HE}/M^\mathrm{WL}$ is biased high for clusters with small WL masses.

The opposite happens at large masses, where the more massive the clusters the rarer. The mass ratio $M^\mathrm{HE}/M^\mathrm{WL}$ is then biased low for clusters with large WL masses.

\subsection{Biased slope}
\label{sec_bias_slope}

The intrinsic scatter in the mass estimate can make the slope of any scaling relation calibrated with either WL or HE masses shallower if the true masses in the selected sample are not uniformly distributed. This is a ripple effect of the Eddington-like bias. Due to measurement errors, the observed variance of the mass proxies is larger than the variance of the true masses. Slope estimators have to correct for this by de-biasing the sample variance \citep{ak+be96}. Here, we are emphasising the similar effect of the intrinsic scatter. Analog treatments, which are often focused on observational errors rather than intrinsic scatter, have already been discussed \citep{an+be12}.

The distribution of the observed mass proxy is smoothed and it has a larger dispersion than the true masses. Due to the finite range, very large (small) measured WL or HE masses likely correspond to smaller (larger) true masses (in arbitrary units), whose observed WL or HE mass were scattered to the tails. If this is not accounted for, the derived slope of the scaling relation is biased toward flatter values.

Let us consider an unbiased (but scattered) mass proxy $X$, i.e., the (logarithm of the) WL or the HE mass, of the (logarithm of the) true mass $Z$,
\beq
\label{eq_sim_1}
X \pm \delta_X = Z \pm \sigma_{X|Z},
\eeq
and a second observable quantity $Y$ we want to calibrate,
\beq
\label{eq_sim_2}
Y \pm \delta_Y = \alpha_{Y|Z} +\beta_{Y|Z} Z \pm \sigma_{Y|Z} .
\eeq
What we usually do is to compare the observable $Y$ to the mass proxy,
\beq
\label{eq_sim_3}
Y \pm \delta_Y =\alpha_{Y|X} +\beta_{Y|X}  X \pm \sigma_{Y|X} .
\eeq
Due to the intrinsic scatter in the mass proxy, $\alpha_{Y|X}$ and $\beta_{Y|X}$ are biased estimates of $\alpha_{Y|Z}$ and $\beta_{Y|Z}$. 

The full scheme outlined before can be formalised through a latent variable model \citep{fe+ba12}. In this approach, $x$ and $y$ are the measured manifest surrogate variables for the unobserved latent true variables $X$ and $Y$, respectively. We refer to \citet[ chapter~7]{fe+ba12} for a review of this and other regression methods for astronomy. Other applications of Bayesian techniques to astronomical contexts can be found in \citet{kel07} and \citet{an+hu12}, and references therein. 

Let us detail our model, which comprises a third latent variable $Z$ together with $X$ and $Y$. The observed values of the mass proxies $X$ and of the observable $Y$ are distributed according to Gaussian distributions centred on the true corresponding values,
\begin{eqnarray}
x_i & \sim & {\cal N}(X_i,\delta_{x,i} ), \label{eq_jags_1}\\
y_i & \sim & {\cal N}(Y_i, \delta_{y,i} ) . \label{eq_jags_2} 
\end{eqnarray}
The true values of the mass proxies and of the observables are scattered with respect the true masses. If the scatters are uncorrelated, the distributions are
\begin{eqnarray}
X_i & \sim & {\cal N}(Z_i,\sigma_{X|Z} ), \label{eq_jags_3} \\
Y_i & \sim & {\cal N}(\alpha_{Y|Z} +\beta_{Y|Z} Z_i, \sigma_{Y|Z}).  \label{eq_jags_4} 
\end{eqnarray}
The distribution of true masses is approximated as a normal function of mean $\mu_Z$ and standard deviation $\sigma_Z$,
\beq
Z_i  \sim {\cal N}(\mu_Z,\sigma_Z ). \label{eq_jags_5}
\eeq
The Gaussian distribution provides a good approximation for signal-selected samples. In fact, at low masses the number of clusters is limited by the selection threshold. At high masses, there are a few clusters because of the steepness of the mass function. The resulting mass distribution is then approximately log-normal for realistic cases, see \citetalias{se+et14_comalit_IV}. Furthermore, the Gaussian function is flexible enough to accommodate for a large range of fairly unimodal distributions \citep{kel07} or even the uniform distribution, which is approximated in case of very large variance.

We chose non-informative priors. For the variances, i.e., the squared scatters, we considered inverse Gamma distribution \citep{an+hu10},
\beq
1/\sigma_{X|Z}^2,\ 1/\sigma_{Y|Z}^2,\ 1/\sigma_{Z}^2 \sim \Gamma(\epsilon,\epsilon),
\eeq
where $\epsilon$ is a small number. In our calculation we took $\epsilon = 10^{-3}$ \citep{an+hu10}. We adopted uniform priors for the intercept $\alpha_{Y|Z}$ and the mean $\mu_Z$. 
\beq
\alpha_{Y|Z},\  \mu_Z  \sim  {\cal U}(-1/\epsilon,1/\epsilon).
\eeq
For the slope $\beta_{Y|Z}$, we assumed a Student's $t_1$ distribution with one degree of freedom, which is equivalent to a uniform prior on the direction angle $\arctan \beta_{Y|Z}$ \citep{an+hu10}, 
\beq
\beta_{Y|Z} \sim  t_1.
\eeq

The unknowns of the regression are the intercept $\alpha_{Y|Z}$, the slope $\beta_{Y|Z}$, and the scatter $\sigma_{Y|Z}$ of the scaling $Y$-$Z$, the scatter $\sigma_{X|Z}$ of the scaling $X$-$Z$, the values of the independent variable $Z_i$ and the parameters describing their distribution $\mu_Z$ and $\sigma_Z$, and the values of the covariate variables $X_i$ and $Y_i$.

We implemented the above Bayesian methodology with JAGS.\footnote{The package JAGS (Just Another Gibbs sampler) by M.~Plummer performs analysis of Bayesian hierarchical models using Markov Chain Monte Carlo simulation. It is publicly available at \url{http://mcmc-jags.sourceforge.net/}.} The scripts used for our analysis can be found at \url{http://pico.bo.astro.it/\textasciitilde sereno/CoMaLit/JAGS/}.

The bias in the estimate of the slope due to the intrinsic scatter in the mass estimate can be studied through a simple simulation. Let us consider a sample of 100 true (logarithmic) masses drawn from a Gaussian distribution with mean $\mu_Z=1.0$ and $\sigma_Z=0.35$. The mass proxies are measured with an observational uncertainty $\delta_x=0.05$. The intrinsic scatter is $\sigma_{X|Z}=0.15$. The proxy $Y$ is linearly related to $Z$ with $\alpha_{Y|Z}=-0.2$ and $\beta_{Y|Z}=1.5$. The scatter around the relation is $\sigma_{Y|Z}=0.2$. The observational uncertainty is fixed to $\delta_{y}=0.05$. We model scatters and errors with normal distributions.

As a first step, we verified that the regression retrieves unbiased parameters when we compare observable $Y$ and true mass $Z$, see Eq.~(\ref{eq_sim_2}). In this simple case we do not need Eqs.~(\ref{eq_jags_1}, and \ref{eq_jags_3}). We found $\alpha_{Y|Z} =-0.22\pm0.07$, $\beta_{Y|Z}=1.49\pm0.07$ and $\sigma_{Y|Z}=0.17\pm0.02$. Regression results are statistically consistent with the input parameters.

We then considered the evolution of $Y$ with the mass proxy $X$, see Eq.~(\ref{eq_sim_3}). In this case, we are not interested in the true values $Z_i$ and we can substitute Eqs.~(\ref{eq_jags_4}, and~\ref{eq_jags_5}) with 
\beq
Y_i  \sim  {\cal N}(\alpha_{Y|X} +\beta_{Y|X} X_i,\sigma_{Y|X}),  \label{eq_jags_6}
\eeq
and
\beq
X_i \sim {\cal N}(\mu_X,\sigma_X ), \label{eq_jags_7}
\eeq
respectively. As before, we do not need Eqs.~(\ref{eq_jags_1}, and~\ref{eq_jags_3}). We found $\alpha_{Y|X} =0.00\pm0.08$, $\beta_{Y|X}=1.25\pm0.07$ and $\sigma_{Y|X}=0.23\pm0.02$. The relation is flatter than the intrinsic one $Y$-$Z$ and the estimated scatter is larger. On turn, the flatter relation causes a higher intercept.

To avoid biases, we have to consider that $X$ is a scattered proxy of the true mass.  Equations~(\ref{eq_sim_1}) and~(\ref{eq_sim_2}) have to be fitted simultaneously and the full scheme in Eqs.~(\ref{eq_jags_1}--\ref{eq_jags_6}) has to be adopted with the additional parameter $\sigma_{X|Z}$. We found $\alpha_{Y|Z} =-0.18\pm0.16$, $\beta_{Y|Z}=1.43\pm0.15$,  $\sigma_{Y|Z}=0.15\pm0.07$ and  $\sigma_{X|Z}=0.11\pm0.05$. Intrinsic parameters are well recovered even though statistical uncertainties are larger.

More details on this statistical model are provided in Appendix~\ref{app_bias}, which also provides some ready-to-use approximate corrections. Correcting $\alpha_{Y|X}$ and $\beta_{Y|X}$ as suggested in Eqs.~(\ref{eq_bias_8}), we found $\alpha_{Y|Z} \sim -0.22$ and $\beta_{Y|Z} \sim 1.48$, in agreement with the input parameters. 

The correction for the biased slope estimated through the widely used BCES method is suggested in Appendix~\ref{app_bces}.

\section{Cluster samples}
\label{sec_samp}

\begin{table*}
\caption{Characteristics of the X-ray and WL samples used in the analysis. $N_\mathrm{cl}$ is the number of clusters in the sample.}
\label{tab_samples}
\begin{tabular}{ l l l l l l l }     
Acronym		 & 	$N_\mathrm{cl}$	&WL instrument& 	WL reference	&	X-ray instrument	&	X-ray reference	& notes\\ 
\hline
RA12  			&	60    		& ---				&	\citet{ras+al12} 	&	--- 			& \citet{ras+al12}	& 	Simulations\\
CCCP-WL  		&	55    		& CFHT			&	\citet{hoe+al12} 	&	---			& 	---	\\
CCCP-HE  		&	50    		& ---				&	---			 	&	Chandra, XMM & \citet{mah+al13}	& 	---	\\
WTG  			&	51    		& Subaru, CFHT	&	\citet{wtg_III_14}	&	--- 			& ---  			& 	---	\\
CLASH-WL		&	20		& Subaru			&	\citet{ume+al14} 	&      --- 			& --- 				& 	---	\\
CLASH-CXO  	&	25    		& ---				&	--- 				&	Chandra 		& \citet{don+al14}  	& 	---	\\
CLASH-XMM  		&	18    		& ---				&	--- 				&	XMM 		& \citet{don+al14}  	& 	---	\\
E10  			&	44    		& ---				&	--- 				&	XMM 		& \citet{ett+al10}  	& 	---	\\
L13  				&	35    		& ---				&	--- 				&	Chandra 		& \citet{lan+al13}  	& 	---	\\
B12  			&	25    		& ---				&	--- 				&	Chandra 		& \citet{bon+al12}  	& 	Additional SZ data from SZA\\
\hline	
\end{tabular}
\end{table*}

We looked in literature for public catalogs compiled in the last few years with either WL or HE masses. The main properties of the samples, which we are going to introduce in the following, are summarised in Table~\ref{tab_samples}.

When quoted mass values were provided with asymmetric errors, we estimated the mean value and the standard deviation as suggested in \citet{dag04}. All the considered masses refer to the fiducial cosmological model. Conversions were performed as described in App.~\ref{app_mass}. A full account of references of WL analyses and standardisation methods can be found in \citetalias{ser14_comalit_III}.

\subsection{Numerical simulations}

\citet[][ RA12]{ras+al12} compared the weak-lensing and X-ray properties of a sample of 20 numerically simulated massive clusters at redshift $z=0.25$. The haloes were the most massive ($M_\mathrm{vir} > 5\times10^{14} M_\odot h^{-1}$) from a set of radiative simulations in a cosmological volume of $1~(\mathrm{Gpc}/h)^3$, evolved in the framework of a WMAP-7 normalised cosmology \citep{fab+al11}. 

Each cluster was later re-simulated at higher resolution and with more complex gas physics. The simulations included: metal-dependent radiative cooling and cooling/heating from a spatially uniform and evolving UV background; a star-formation model where a hot ionised phase coexists in pressure equilibrium with a cold phase, which is the reservoir for star formation; a description of metal enrichment from different stellar populations; the effect of supernovae feedback through galactic winds. 

The clusters were finally processed to generate optical and X-ray mock observations along three orthogonal projections. The final sample consists of 60 cluster realisations. WL and HE masses are estimated within $r_{500}^\mathrm{Tr}$, the over-density radius corresponding to the true mass.

\subsection{Canadian Cluster Comparison Project }

The Canadian Cluster Comparison Project \citep[CCCP,][]{mah+al13} assembled a sample of 50 rich clusters of galaxies in the redshift range $0.15 < z < 0.55$. All of the clusters were observable from the Canada-France-Hawaii Telescope (CFHT), which restricts the sample to systems at $-15\deg <$ declination $< 65\deg$. Most of them were selected to have an ASCA (Advanced Satellite for Cosmology and Astrophysics) temperature $k_\mathrm{B} T_X > 3~\mathrm{keV}$. X-ray properties were measured either with Chandra or XMM-Newton. Weak lensing studies for 5 additional clusters without X-ray analyses can be found in \citet{hoe+al12}.

Lensing masses were determined with aperture statistics \citep{hoe+al12}. This approach relies on shear measurements at large radii and reduces the effects of contamination by cluster members. The 3D masses were computed from the model-independent 2D aperture masses with a de-projection method based on a NFW density profile. 

\citet{mah+al13} performed the X-ray analysis of the sample using both Chandra and XMM observations. They found that due to temperature discrepancies, the XMM cluster masses were systematically $\sim$ 15 per cent smaller than Chandra masses. In order to combine the data, \citet{mah+al13} down-weighted the high-energy effective area of Chandra. X-ray quantities were estimated either within $r_{500}^\mathrm{WL}$, the radius evaluated from the weak-lensing mass measurement, or $r_{500}^\mathrm{HE}$, as evaluated from the mass estimate assuming hydrostatic equilibrium.\footnote{Values of $M_{500}^\mathrm{WL}$ and $M_{500}^\mathrm{HE}$ for the CCCP sample are publicly available at \url{http://sfstar.sfsu.edu/cccp}.}

\citet{mah+al13} identified a subsample of 20 cool core systems with core entropy at 20~kpc smaller than 70~$\mathrm{keV~cm^2}$ and 8 systems with low offsets between the brightest cluster galaxy (BCG) and the X-ray surface brightness peak, $D_\mathrm{BCG}<10~\mathrm{kpc}$.

\subsection{Cluster Lensing And Supernova survey with Hubble}

The CLASH programme \citep{pos+al12} has been mapping the matter distribution of 25 rich clusters drawn largely from the Abell and MAssive Cluster Survey \citep[MACS,][]{ebe+al10} cluster catalogs. \citet{ume+al14} performed a joint shear-and-magnification weak-lensing analysis of a sub-sample of 16 X-ray regular and 4 high-magnification galaxy clusters in the redshift range $0.19 \la z \la 0.69$.  A complementary analysis exploiting strong lensing data was presented in \citet{mer+al14}. 

To make the comparison with the other data samples easier, we will use the mass estimates in \citet{ume+al14}, whose methodology exploits only the weak-lensing regime whereas results in \citet{mer+al14} strongly rely on information from the inner regions. Mass estimates in \citet{ume+al14} were based on joint weak lensing shear plus magnification measurements based on ground-based wide-field Subaru data. On the other hand, the analysis in \citet{mer+al14} combined the Subaru shear profile with weak-lensing constraints from the Hubble Space Telescope (HST) in the intermediate regime and strong lensing constraints from HST. 

All of the CLASH clusters have been observed with the Chandra satellite \citep{pos+al12}. A subsample of 18 clusters was targeted by XMM too. The X-ray analysis was presented in \citet{don+al14}, which computed HE masses and gas fractions.

Based on Chandra data, \citet{don+al14} identified 10 clusters (9 of them with WL mass) with a strong cool core, i.e., with an excess core entropy smaller than 30~$\mathrm{keV~cm^2}$.

\subsection{Weighing the Giants}

The Weighing the Giants \citep[WTG,][]{wtg_I_14} program targeted 51 X-ray luminous clusters from the MACS and the Brightest Cluster Survey \citep[BCS,][]{ebe+al00}. The clusters span a large range in redshift ($0.15 \la z \la 0.7$) and dynamical state. Seven clusters are classified as relaxed \citep{wtg_I_14}.
 
The values of the scale radius and the concentration are provided in \citet[ table 4]{wtg_III_14}. We derived $M_{500}^\mathrm{WL}$ and $r_{500}^\mathrm{WL}$ using the NFW density profile adopted in the WTG analysis.

\subsection{X-ray samples}

\citet[][ E10]{ett+al10} studied a sample of 44 X-ray luminous galaxy clusters observed with XMM-Newton in the redshift range $0.1\la z \la0.3$. They applied two different techniques (the backward `method 1', which we take as the reference method, and the forward `method 2') to recover the gas and the dark mass properties, described with a NFW profile. Clusters were classified according to their core properties. E10 identified a subsample of 16 low-entropy-core systems, which represent the prototype of relaxed clusters with a well defined cool core at low entropy.

\citet[][ L13]{lan+al13} presented Chandra X-ray measurements of the hydrostatic mass and of the gas mass fraction out to $r_{500}$ for the complete sample of the 35 most luminous clusters from the BCS and its extension at redshift $0.15\la z \la 0.30$. The clusters span a large range of dynamical states. The data were analysed using two independent pipelines and two different models for the gas density and temperature, the `Polytropic' (which we take as our reference case) or the `Vikhlinin' model. 

\citet[][ B12]{bon+al12} derived the hydrostatic masses and the pressure profile of a sample of 25 massive relaxed galaxy clusters with a simultaneous analysis of SZ data from the Sunyaev-Zel'dovich array (SZA) and archival Chandra observations. 




\section{Mass comparison}
\label{sec_mass_comp}

Even though in principle WL and X-ray masses could be unambiguously determined from a given set of observations, calibration issues and hidden systematics make these measurements very difficult.

In this section we compare either WL or HE masses from different catalogs. It is nowadays customary to quote masses within a given over-density and to derive scaling relations in terms of them. These masses can be related to the virial mass and most cluster properties are expected to be self-similar if rescaled by their value at $r_\Delta$. To limit extrapolation of published results, we then considered the masses within $r_{500}$, rather than extrapolating the results up to a fixed length. 

On the other hand, the relationship between $M_\Delta$ and $r_\Delta$ exacerbates problems connected to aperture differences, which complicate the comparison between different samples. Since the total mass within a fixed radius scales nearly linearly with the radius, differences in mass within a given over-density are inflated by $\sim 100/3$ per cent with respect to differences within a fixed physical radius.

Differences among properties measured within a fixed length are not inflated but they refer to physically different regions in differently sized clusters. A promising alternative is to express the results in terms of the circular velocity $v_\mathrm{Circ}^2 = G M(r)^2/r$. In fact, the circular velocity is almost independent of cosmology and it is nearly unaffected by aperture problems \citep{don+al14}. Within a given over-density radius, the velocity scales with mass as $v_\Delta \propto M_\Delta^{2/3}$. Quoted results for central estimate and scatter of $\Delta \ln M_{500}$, as well as fractional changes, can be translated in analogue results for $\Delta \ln v_{500}$ by simply multiplying by the factor 2/3.

To compare different samples we considered the (natural) logarithm of mass ratios \citep{roz+al14}. The central estimate and the scatter were computed as bi-weight estimators of the distribution. Uncertainties were estimated with bootstrap resampling with replacement. The main advantage in using logarithms is that their difference is (anti-)symmetric. This solves the problem affecting those estimators of ratio which are not symmetric with respect to the exchange of numerator and denominator.

Quoted errors in compiled catalogs may account for different sources of statistical and systematic uncertainties. Furthermore, it can be argued that the published uncertainties are unable to account for the actual variance seen in sample pairs \citep{roz+al14}. We then conservatively performed unweighted analyses.

\subsection{WL masses}

\begin{table}
\caption{Comparison of WL masses from independent analyses. We quote the mean $\ln$ differences in mass for sample pairs. Entries are in the format: $(N_\mathrm{cl}), \mu(\pm \delta\mu)\pm \sigma(\pm\delta \sigma)$, where $N_\mathrm{cl}$ is the number of clusters in common between the samples, $\mu$ is the central estimate of the difference in natural logarithm $\ln ( M_{500}^{\mathrm{row}}/M_{500}^\mathrm{col})$, with associated uncertainty $\delta \mu$; $\sigma$ is the dispersion with associated uncertainty $\delta \sigma$. $M_{500}^{\mathrm{row}}$ ($M_{500}^\mathrm{col}$) refers to the sample indicated in the corresponding row (column). Quoted values are the bi-weight estimators.}
\label{tab_comp_MWL}
\centering
\begin{tabular}[c]{ l    c   c }     
 				&	CLASH-WL			&	WTG	\\ 
	 \hline
  			  	&	$(6)$ 				&  	$(17)$  			\\
CCCP-WL  		  	&	$-0.45(\pm0.12)$ 		&	$-0.31(\pm0.05)$  	\\
			  	&	$\pm 0.25(\pm 0.10)$ 	&  	$ \pm 0.21(\pm 0.09)$  \\
	\hline
			 	& 	-- 					& 	$(17)$  			\\
CLASH-WL 		 	& 	-- 					& 	$0.01(\pm0.10)$  	\\
			 	& 	-- 					& 	$\pm 0.37(\pm 0.06)$  \\
	\hline	
\end{tabular}
\end{table}

\begin{figure}
       \resizebox{\hsize}{!}{\includegraphics{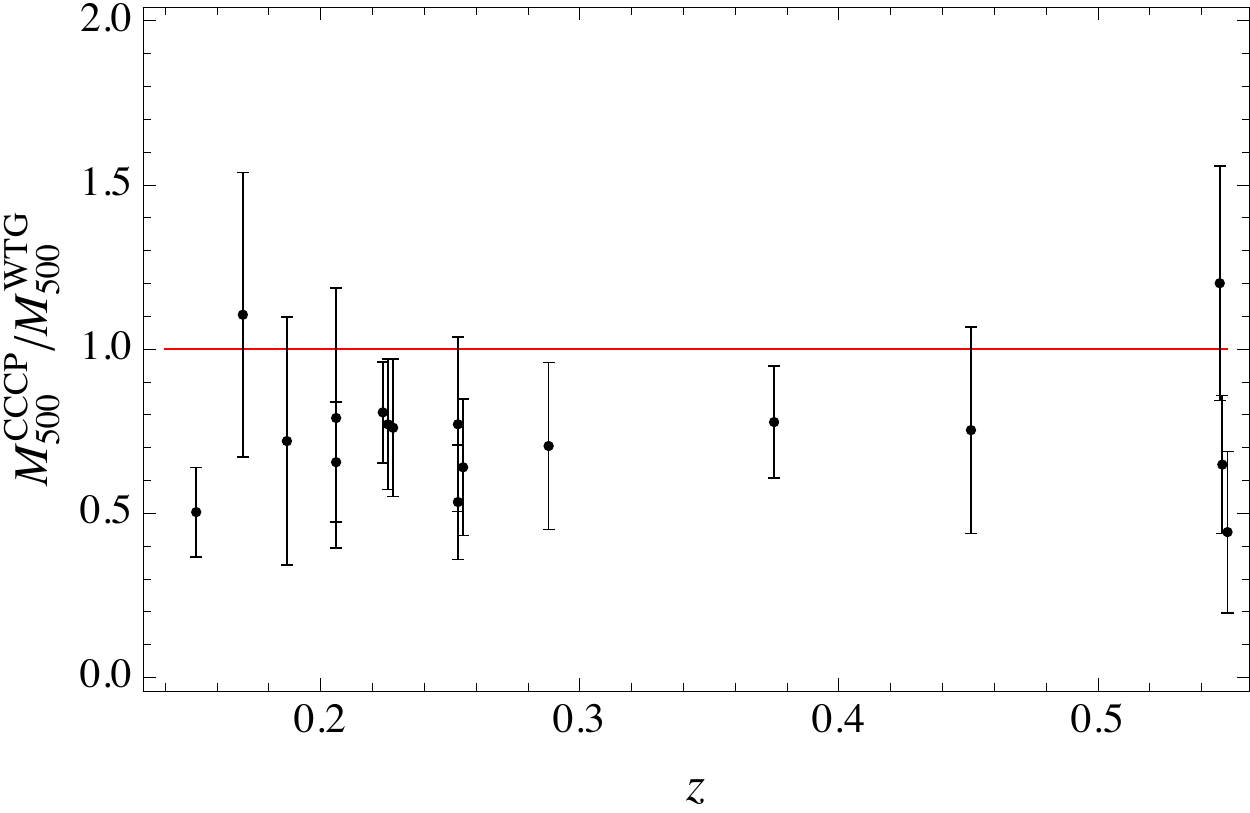}}
       \caption{Ratio of WL masses estimated by CCCP and WTG as a function of redshift.}
	\label{fig_M500_CCCP_vs_WTG}
\end{figure}

In principle, WL masses could be determined to an accuracy of $\la 8$ per cent \citep{wtg_I_14,ume+al14}, but differences between masses reported by distinct groups are off by $\sim$ 20-50 per cent \citep{wtg_III_14,ume+al14}. On the other hand, comparisons show that mass measurements correlate quite tightly \citep{wtg_III_14}.

The CCCP and the WTG samples share 17 clusters, see Table~\ref{tab_comp_MWL}. $M_{500}$ from CCCP are smaller by $\sim 30$ per cent with a scatter of $\sim 20$ per cent. This difference is way larger than the claimed mass calibration uncertainty and highlights the difficulties connected to unbiased calibrations in WL measurements. We found no trend with redshift, see Fig.~\ref{fig_M500_CCCP_vs_WTG}. 

The CCCP masses are notably underestimated with respect to the CLASH clusters too. On the other hand, the agreement between the WTG and the CLASH results is substantial, even though the scatter in the mass ratios is quite large. Usually the two mass estimates of a given cluster from WTG and CLASH coincide within $\la 30$ per cent of the combined error.

The scatters in the mass ratio are of order of 20-40 per cent and are consistent with the quoted statistical uncertainties on the WL mass estimates, which are of the order of 20 per cent or larger. In fact, if statistical errors are properly estimated, the combined scatter in the mass ratios should be approximately given by the quadratic sum of the typical errors of the two considered samples.

\subsection{X-ray masses}

\begin{table}
\caption{Comparison of HE masses from independent methods but from the same data-sets. E10-M1 and E10-M2 denote the two methods used in \citet{ett+al10}. L13-Vick and L13-Poly denote the two methods used in \citet{lan+al13}. $N_\mathrm{cl}$ (col. 3) is the number of clusters in common between the samples listed in cols. 1 and 2. $M_{500}^{(1)}/M_{500}^{/2)}$ (col. 4) is the central estimate of the ratio between masses in the two samples; $\sigma$ is the dispersion. Quoted values are the bi-weight estimators.}
\label{tab_comp1_MHE}
\centering
\begin{tabular}[c]{l l c  r@{$\,\pm\,$}l r@{$\,\pm\,$}l}
        \hline
        \noalign{\smallskip}
	Sample 1 	& Sample 2 	& $N_\mathrm{cl}$	&\multicolumn{2}{c}{$M_{500}^{(1)}/M_{500}^{(2)}$} & \multicolumn{2}{c}{$\sigma$}  \\
        \noalign{\smallskip}
        \hline
	E10-M2		&	E10-M1			&	44	& 0.98	&	0.05	&	0.25	&	0.03	\\
	L13-Vick		&	L13-Poly			&	35	& 0.99	&	0.04	&	0.23	&	0.03	\\
	\hline
	\end{tabular}
\end{table}

\begin{table*}
\caption{Comparison of HE masses from independent analyses. For the CCCP-HE sample, we considered masses within $r^\mathrm{HE}_{500}$.  We quote the mean $\ln$ differences in mass for sample pairs. Entries are in the format: $(N_\mathrm{cl}), \mu(\pm \delta\mu)\pm \sigma(\pm\delta \sigma)$, where $N_\mathrm{cl}$ is the number of clusters in common between the samples, $\mu$ is the central estimate of the difference in natural logarithm $\ln ( M_{500}^{\mathrm{row}}/M_{500}^\mathrm{col})$, with associated uncertainty $\delta \mu$; $\sigma$ is the dispersion with associated uncertainty $\delta \sigma$. $M_{500}^{\mathrm{row}}$ ($M_{500}^\mathrm{col}$) refers to the sample indicated in the corresponding row (column). Quoted values are the bi-weight estimators.}
\label{tab_comp_2_MHE}
\begin{tabular}[c]{|cccccc}     
				&	CCCP-HE				&	CLASH-XMM			&	CLASH-CXO	&	L13				&	B12	\\ 
\hline
  				&	$(11)$ 				&	$(3)$ 			&  	$(3)$ 				&	$(11)$ 			& 	$(6)$ \\
E10  			&	$0.22(\pm0.08)$ 		&	$\sim0.17$		&  	$\sim-0.15$ 			&	$0.35(\pm0.14)$ 	& 	$0.25(\pm0.10)$ \\
  				&	$\pm0.28(\pm0.11)$ 	&	$\pm(\sim)0.06$	&  	$\pm(\sim)0.28$ 		&	$\pm0.30(\pm0.09)$ & 	$\pm0.19(\pm0.08)$ \\
\hline
 				& 						& 	$(5)$  			& 	$(6)$ 				& 	$(18)$  			&  	$(5)$ 			\\
CCCP-HE 		& 	--- 					& 	$0.03(\pm0.21)$	& 	$-0.38(\pm0.14)$ 		& 	$0.12(\pm0.07)$ 	&  	$0.24(\pm0.22)$ \\
 				& 	 					& 	$\pm0.29(\pm0.16)$	& 	$\pm0.34(\pm 0.21)$ 	& 	$\pm0.33(\pm0.14)$	&  	$\pm0.35(\pm0.17)$ \\
\hline
				&						&		 			&	 (18) 				&	$(2)$			& 	$(10)$ \\
CLASH-XMM		&	---					&	---			 	&	$-0.38(\pm0.09)$  		&	$\sim-0.05$		& 	$0.14(\pm0.15)$ \\
				&						&					&	 $\pm0.35(\pm0.10)$ 	&	$\pm(\sim)0.18$	& 	$\pm0.46(\pm0.30)$ \\
\hline
				&						&		 			&						&	$(4)$ 			& 	$(12)$ \\
CLASH-CXO	&	---					&	--- 	 			&	---					&	$\sim0.31$ 		& 	$0.45(\pm0.14)$ \\
				&						&	 				&						&	$\pm(\sim)0.01$ 	& 	$\pm0.37(\pm0.13)$ \\
\hline
				&						&					&						& 	 				&	$(4)$ \\
L13				&	---					&	---				&	 ---					& 	--- 				&	$\sim0.03$ \\
				&						&					&	 					& 					&	$\pm(\sim)0.06$ \\
\hline			
\end{tabular}
\end{table*}

\begin{figure}
\begin{tabular}{c}
\includegraphics[width=8.5cm]{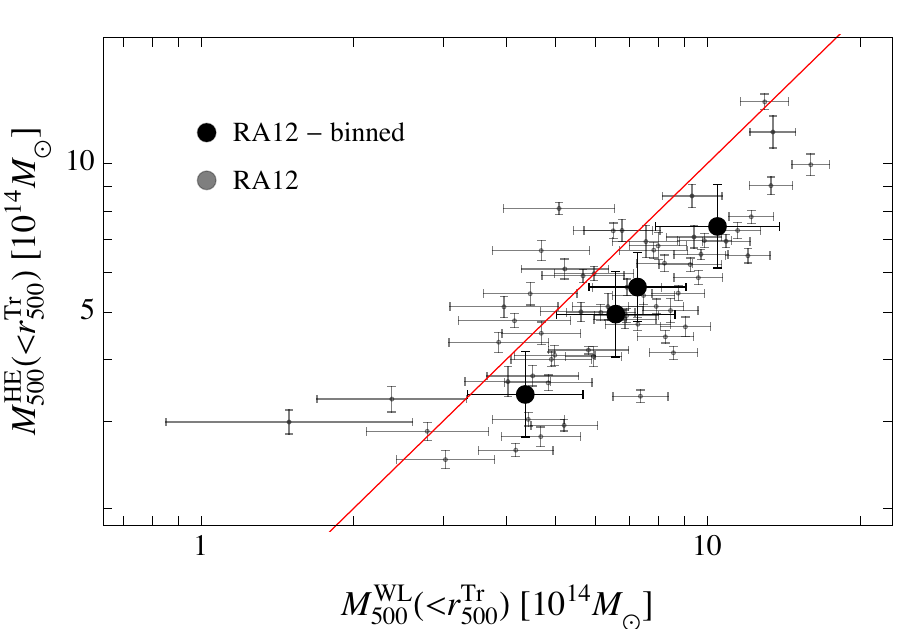} \\
\noalign{\smallskip}  
\includegraphics[width=8.5cm]{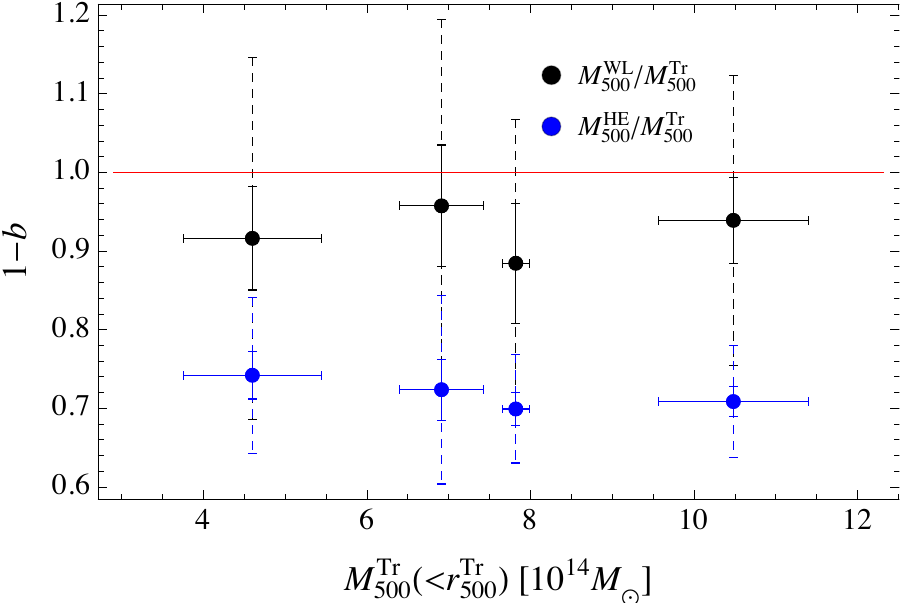} \end{tabular}
\caption{Masses in the RA12 sample. {\it Top panel}: WL mass vs. HE mass. Clusters are grouped in four bins in true mass (black points). The red line is the bisector $M_{500}^\mathrm{HE}=M_{500}^\mathrm{WL}$. {\it Lower panel}: bias of the proxy as a function of the true mass. Clusters are grouped in four bins in true mass. Black (blue) points correspond to the bias of the WL (HE) mass. The solid error-bars denote the 1-$\sigma$ uncertainties for the the central estimate. The dashed error-bars denote the dispersion. All masses are computed within $r_{500}^\mathrm{Tr}$.}
\label{fig_RA12_MWL_MHE_bin_MTr}
\end{figure}

X-ray properties of galaxy clusters reported by competing groups may reach discrepancies of 50 per cent \citep{roz+al14}. Here, we consider the off-set and the scatter in the estimate of HE masses. Discrepancies may stem from either differences in the considered data sets (to the larger extent if taken with different instruments), or from not consistent data reduction pipelines, or from different techniques to recover the mass. 

This last issue can be quantified by comparing mass estimates obtained from the same data-sets but with different methodologies. This is the case of the analyses in either E10 or L13, for which we could compare the scatter in the mass estimate due to the different modelling, see Table~\ref{tab_comp1_MHE}. 

The typical statistical error in a HE mass estimate is of the order of $\sim 15$ per cent. The observed scatter in the mass ratios is then consistent with the propagation of this error. This comparison suggests that mass estimates are not biased due to different techniques, whose associated variance is negligible with respect to the statistical uncertainty.

Larger variations are mainly related to different data-sets, see Table~\ref{tab_comp_2_MHE}. Discrepancies of order of $\ga 30$ per cent may be in place. This is the case for results based on Chandra (CLASH-CXO, B12, L13) versus XMM analyses (E10, CLASH-XMM), whose temperature estimates may disagree at large radii \citep{don+al14}. 

Each method/analysis may systematically either under o over-estimate the cluster mass. X-ray masses in the CLASH sample based on Chandra (XMM) data are systematically larger (smaller) than other estimates. On the other hand, masses from B12 and L13 are lower than other samples.

A significant role can be played by additional data-sets exploited in the analysis. The inclusion of SZ data, which are more sensitive to the outer regions, might lower the mass values in B12.

The large differences in estimated masses and the large scatters suggest that quoted formal statistical uncertainties in HE masses, usually of the order of $\sim$10-15 per cent, might be under-estimated.

\section{Regression results}
\label{sec_resu}

We measured biases and intrinsic scatters of WL and HE masses through the statistical model detailed in Sec.~\ref{sec_bias}. To simplify the analysis, we assumed that the lensing and the hydrostatic masses scale linearly with the true mass, $\beta_\mathrm{WL}=1$ and $\beta_\mathrm{HE}=1$. In relation to the notation in Sec.~\ref{sec_bias_slope}, the logarithm of the WL mass ($\ln M^\mathrm{WL}$), of the HE mass ($\ln M^\mathrm{HE}$), and of the true mass ($\ln M^\mathrm{Tr}$) can be identified with $X$, $Y$, and $Z$, respectively.

The true masses are known only in simulations. For observed samples, we can estimate only the relative bias between WL and HE masses and we fixed $\alpha_\mathrm{WL}=0$. The effective bias $M^\mathrm{HE} /M^\mathrm{Tr}$ ($M^\mathrm{WL} /M^\mathrm{Tr}$) can be defined as $\exp [\alpha_\mathrm{HE}]$ ($\exp [\alpha_\mathrm{WL}]$). The relative bias $M^\mathrm{HE}/M^\mathrm{WL}$ can be defined as $\exp (\alpha_\mathrm{HE}-\alpha_\mathrm{WL})$. Bias and scatter are largely uncorrelated. We tested that the estimates of scatter and relative bias do not change whether we consider $\alpha_\mathrm{HE}=0$ rather than $\alpha_\mathrm{WL}=0$.

The intrinsic distribution of the independent variable, $\ln M^\mathrm{Tr}$, was approximated with a Gaussian function, see Sec.~\ref{sec_bias_slope}. We tested that results based on more complex distributions, such as mixtures of Gaussian functions \citepalias{ser+al14_comalit_II}, were indistinguishable from the simplest case.

\subsection{Simulated sample}

\begin{figure*}
\begin{tabular}{cc}
 \includegraphics[width=8.5cm]{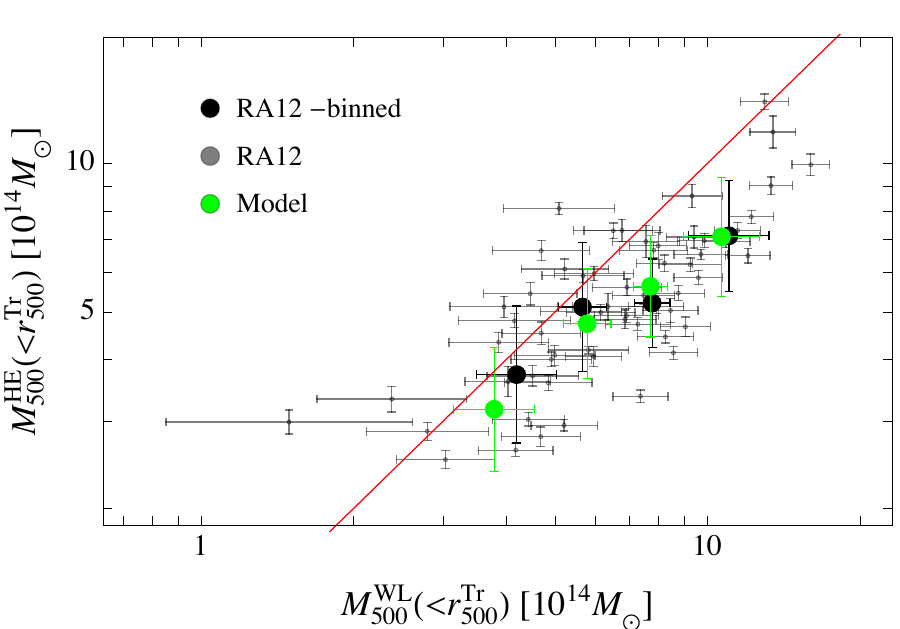}  & \includegraphics[width=8.5cm]{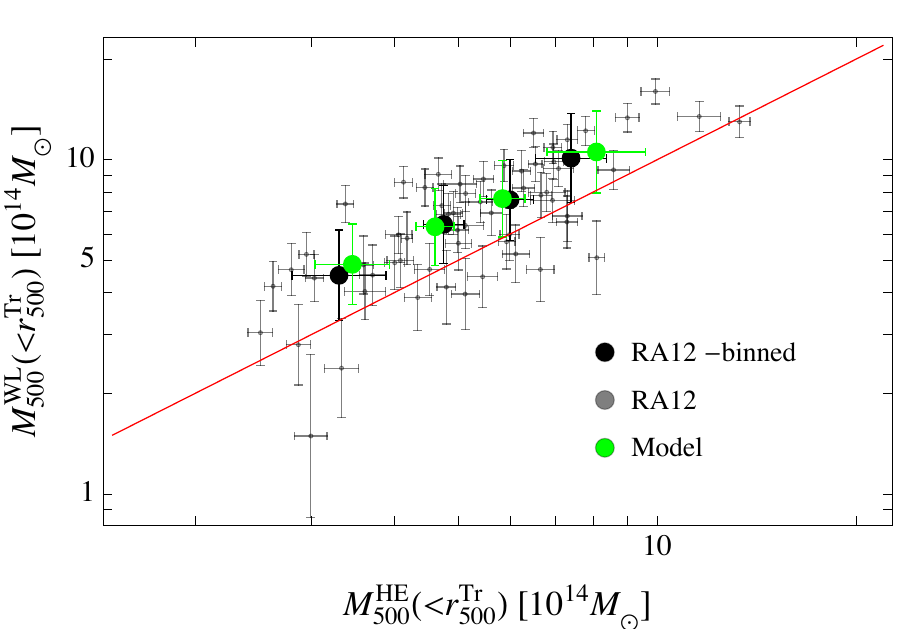} \\
\noalign{\smallskip}  
\includegraphics[width=8.5cm]{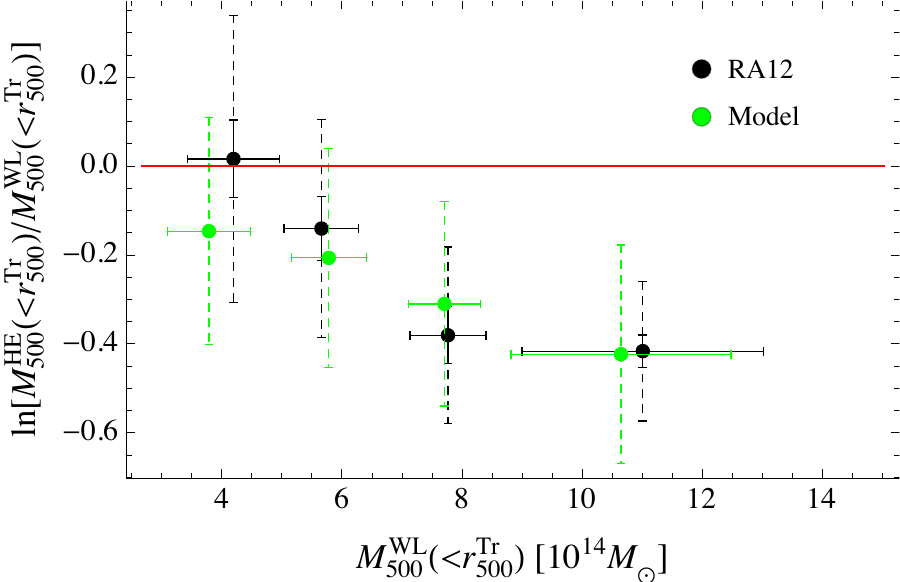} &	\includegraphics[width=8.5cm]{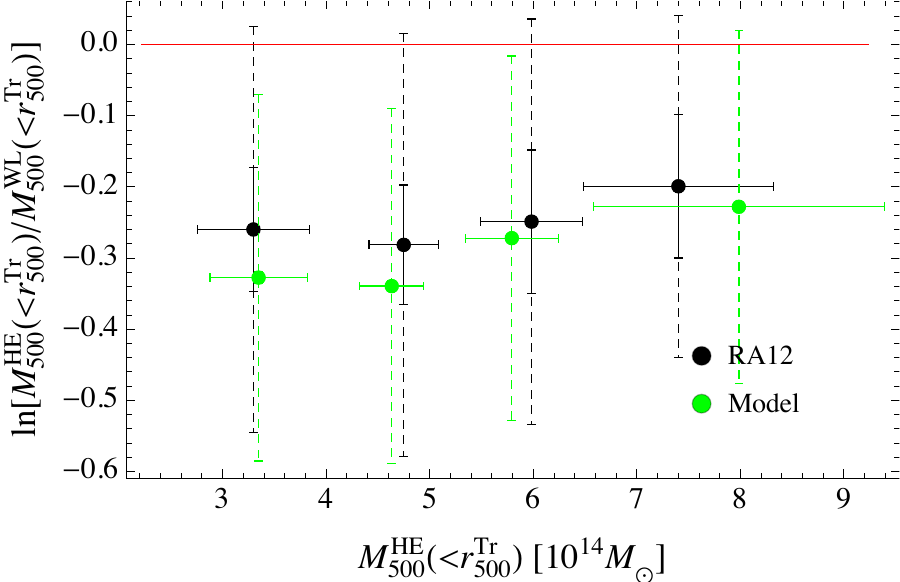}
\end{tabular}
\caption{Comparison of WL and HE masses for the RA12 sample. Masses are measured within $r_{500}^\mathrm{Tr}$, the over-density radius related to the true mass. In the left (right) panels, clusters are grouped in 4 bins according to their measured WL (HE) mass. Green points mark the predictions based on the analytical model discussed in the paper. {\it Top left panel}: $M_{500}^\mathrm{HE}$ as a function of $M_{500}^\mathrm{WL}$. Black (green) points mark clusters (analytical predictions) binned in $M_{500}^\mathrm{WL}$. Errors bars for the binned points are computed as the dispersion around the central value. {\it Bottom left panel}: logarithm of $M_{500}^\mathrm{HE}/M_{500}^\mathrm{WL}$ as a function of $M_{500}^\mathrm{WL}$. The solid error-bars denote the 1-$\sigma$ uncertainties for the the central estimate. The dashed error-bars denote the dispersion. {\it Top right panel}: $M_{500}^\mathrm{WL}$ as a function of $M_{500}^\mathrm{HE}$. Black points marks clusters binned in $M_{500}^\mathrm{HE}$. {\it Bottom right panel}: logarithm of $M_{500}^\mathrm{HE}/M_{500}^\mathrm{WL}$ as a function of $M_{500}^\mathrm{HE}$.}
\label{fig_RA12_MWL_MHE}
\end{figure*}

\begin{table}
\caption{Intrinsic scatter and bias of mass proxies $M^\mathrm{WL}_{500}$ and $M_{500}^\mathrm{HE}$ for the RA12 clusters.}
\label{tab_scat_RA12}
\centering
\begin{tabular}[c]{l   r@{$\,\pm\,$}l r@{$\,\pm\,$}l}
        \hline
        \noalign{\smallskip}
	$M^\mathrm{Proxy}$ 	& \multicolumn{2}{c}{$M_{500}^\mathrm{Proxy}/M_{500}^\mathrm{Tr}$} & \multicolumn{2}{c}{$\sigma$}  \\
        \noalign{\smallskip}
        \hline
	$M_{500}^\mathrm{WL}$		&	0.95	&	0.03	&	0.13	&	0.03	\\
	\hline
	$M_{500}^\mathrm{HE}$		&	0.71	&	0.01	&	0.11	&	0.02	\\
	\hline
	\end{tabular}
\end{table}

As a first step, we analysed the simulated sample from RA12. In the realm of simulations, we know the true masses of the clusters. We can exploit this information to compute the bias and the intrinsic scatter of each mass proxy separately by direct comparison with the true mass. WL and HE masses can be compared to true masses autonomously. Regression results are in agreement with the original analysis in \citet{ras+al12} and are summarised in Table~\ref{tab_scat_RA12}. Note that differently from \citet{ras+al12}, we estimated the intrinsic rather than the total scatter and we focused on logarithmic variables.

The level of bias for each proxy is approximately constant with respect to the true mass, see Fig.~\ref{fig_RA12_MWL_MHE_bin_MTr}. 

The intrinsic scatter plays an important role when we analyse the bias as a function of the mass proxy, see Fig.~\ref{fig_RA12_MWL_MHE}. The decreasing trend of the ratio $M^\mathrm{HE}_{500}/M^\mathrm{WL}_{500}$  as a function of $M^\mathrm{WL}_{500}$ is an effect of the scatter of $M^\mathrm{WL}_{500}$ around the true mass, see Sec.~\ref{sec_bias_edd}.

Due to the combined action of selection effect and intrinsic scatters, at small (large) values of WL masses, $M^\mathrm{WL}_{500}$ is biased low (high) with respect to the true mass whereas $M^\mathrm{HE}_{500}$ is unbiased. As a consequence, the ratio $M^\mathrm{HE}_{500}/M^\mathrm{WL}_{500}$ decreases with $M^\mathrm{WL}_{500}$. At intermediate values, clusters can be scattered into a given range in $M^\mathrm{WL}_{500}$ from either above or below, and the ratio $M^\mathrm{HE}_{500}/M^\mathrm{WL}_{500}$ is not biased. The larger the scatter, the steeper the trend in the mass ratio as a function of the mass proxy. The scatter also flattens the $M^\mathrm{HE}_{500}$-$M^\mathrm{WL}_{500}$ relation towards larger values of $M^\mathrm{WL}_{500}$. For similar reasons and being $M^\mathrm{HE}_{500}$ at the denominator, the ratio $M^\mathrm{WL}_{500}/M^\mathrm{HE}_{500}$ increases as a function of $M^\mathrm{HE}_{500}$.

The statistical model summarised in Eqs.~(\ref{eq_proxy_WL}, and~\ref{eq_proxy_HE}) and detailed in Sec.~\ref{sec_bias_slope} accounts for these effects and can be validated by comparing its predictions to the RA12 sample. We generated a sample of simulated clusters whose true masses are drawn from a Gaussian distribution. The corresponding measured WL and HE masses were simulated assuming intrinsic scatters and observational uncertainties as measured in the RA12 sample. This model successfully reproduces the trends in the observed mass ratio, see Fig.~\ref{fig_RA12_MWL_MHE}. 

As a second step, we tested the regression algorithm with the mock observations of the simulated RA12 sample, see Table~\ref{tab_scatter}. Differently from the first step, we used the manifest estimated values of the mass proxies but we did not exploit the information on the latent true mass. We could then not calibrate the bias in either $M^\mathrm{WL}_{500}$ or $M^\mathrm{WL}_{500}$ in an absolute way, but we had to normalise one bias relatively to the other one. We assumed $\alpha_\mathrm{WL}=0$, i.e., we measured $M_{500}^\mathrm{HE}/M_{500}^\mathrm{Tr}$ in units of $M_{500}^\mathrm{WL}/M_{500}^\mathrm{Tr}$. The level of relative bias and the intrinsic scatters are recovered within the statistical uncertainties.

\subsection{Observed samples}

\begin{table*}
\caption{
Biases and intrinsic scatters of the WL and HE masses. Col. 1: sample; col. 2: number of clusters in the sample, $N_\mathrm{cl}$; cols. 3, 4: radius within which the WL lensing and the HE mass were computed, respectively; col. 5: effective ratio between the true mass and the WL mass; the WL mass is assumed to be an unbiased proxy; col. 6: intrinsic scatter of $\ln M_{500}^\mathrm{WL}/M_{500}^\mathrm{Tr}$; col. 7: effective ratio $M_{500}^\mathrm{HE}/M_{500}^\mathrm{WL}$; col.~8: intrinsic scatter (as in col.~6 but for HE masses). Quoted values are bi-weight estimators of the posterior probability distribution.}
\label{tab_scatter}
\begin{tabular}[c]{l l	c c c r@{$\,\pm\,$}l r@{$\,\pm\,$}l r@{$\,\pm\,$}l}
        \hline
	Sample				& $N_\mathrm{cl}$		& $r^\mathrm{WL}$	&  $r^\mathrm{HE}$ 	& $M_{500}^\mathrm{WL}/M_{500}^\mathrm{Tr}$ & \multicolumn{2}{c}{$\sigma_\mathrm{WL}$} 	& \multicolumn{2}{c}{$M_{500}^\mathrm{HE}/M_{500}^\mathrm{WL}$}	& \multicolumn{2}{c}{$\sigma_\mathrm{HE}$}  \\
        \hline
	RA12					&	60	&	$r_{500}^\mathrm{Tr}$	&	$r_{500}^\mathrm{Tr}$	&	[1]	&	0.14	&	0.04	&	0.75	&	0.03	&	0.13	&	0.04	\\
	\hline
	CCCP					&	50	&	$r_{500}^\mathrm{WL}$	&	$r_{500}^\mathrm{WL}$	&	[1]	&	0.14	&	0.06	&	0.85	&	0.05	&	0.24	&	0.07	\\
	 \noalign{\smallskip}
	CCCP-Cool Core			&	16	&	$r_{500}^\mathrm{WL}$	&	$r_{500}^\mathrm{WL}$	&	[1]	&	0.18	&	0.10	&	0.93	&	0.11	&	0.24	&	0.12	\\
	 \noalign{\smallskip}
	CCCP-Low Offset			&	20	&	$r_{500}^\mathrm{WL}$	&	$r_{500}^\mathrm{WL}$	&	[1]	&	0.18	&	0.10	&	0.82	&	0.09	&	0.30	&	0.11	\\
	 \noalign{\smallskip}
	CCCP					&	50	&	$r_{500}^\mathrm{WL}$	&	$r_{500}^\mathrm{HE}$	&	[1]	&	0.20	&	0.09	&	0.81	&	0.07	&	0.45	&	0.07	\\
	\hline
	CLASH-CXO				&	20	&	$r_{500}^\mathrm{WL}$	&	$r_{500}^\mathrm{HE}$	&	[1]	&	0.17	&	0.09	&	0.78	&	0.09	&	0.34	&	0.12	\\
	\noalign{\smallskip}
	CLASH-CXO-Cool Core		&	9	&	$r_{500}^\mathrm{WL}$	&	$r_{500}^\mathrm{HE}$	&	[1]	&	0.22	&	0.14	&	0.77	&	0.14	&	0.31	&	0.17	\\
	\noalign{\smallskip}
	CLASH-CXO-Low Offset		&	8	&	$r_{500}^\mathrm{WL}$	&	$r_{500}^\mathrm{HE}$	&	[1]	&	0.31	&	0.17	&	0.70	&	0.15	&	0.34	&	0.17	\\
	\hline
	\noalign{\smallskip}
	CLASH-XMM				&	16	&	$r_{500}^\mathrm{WL}$	&	$r_{500}^\mathrm{HE}$	&	[1]	&	0.17	&	0.10	&	0.56	&	0.08	&	0.45	&	0.14	\\
	\hline
	WTG-L13					&	14	&	$r_{500}^\mathrm{WL}$	&	$r_{500}^\mathrm{HE}$	&	[1]	&	0.32	&	0.14	&	0.64	&	0.09	&	0.16	&	0.08	\\
	\noalign{\smallskip}
	WTG-B12					&	14	&	$r_{500}^\mathrm{WL}$	&	$r_{500}^\mathrm{HE}$	&	[1]	&	0.19	&	0.12	&	0.47	&	0.07	&	0.34	&	0.15	\\
	\hline
	\end{tabular}
\end{table*}

\begin{figure*}
\begin{tabular}{rrr}
\includegraphics[width=5cm]{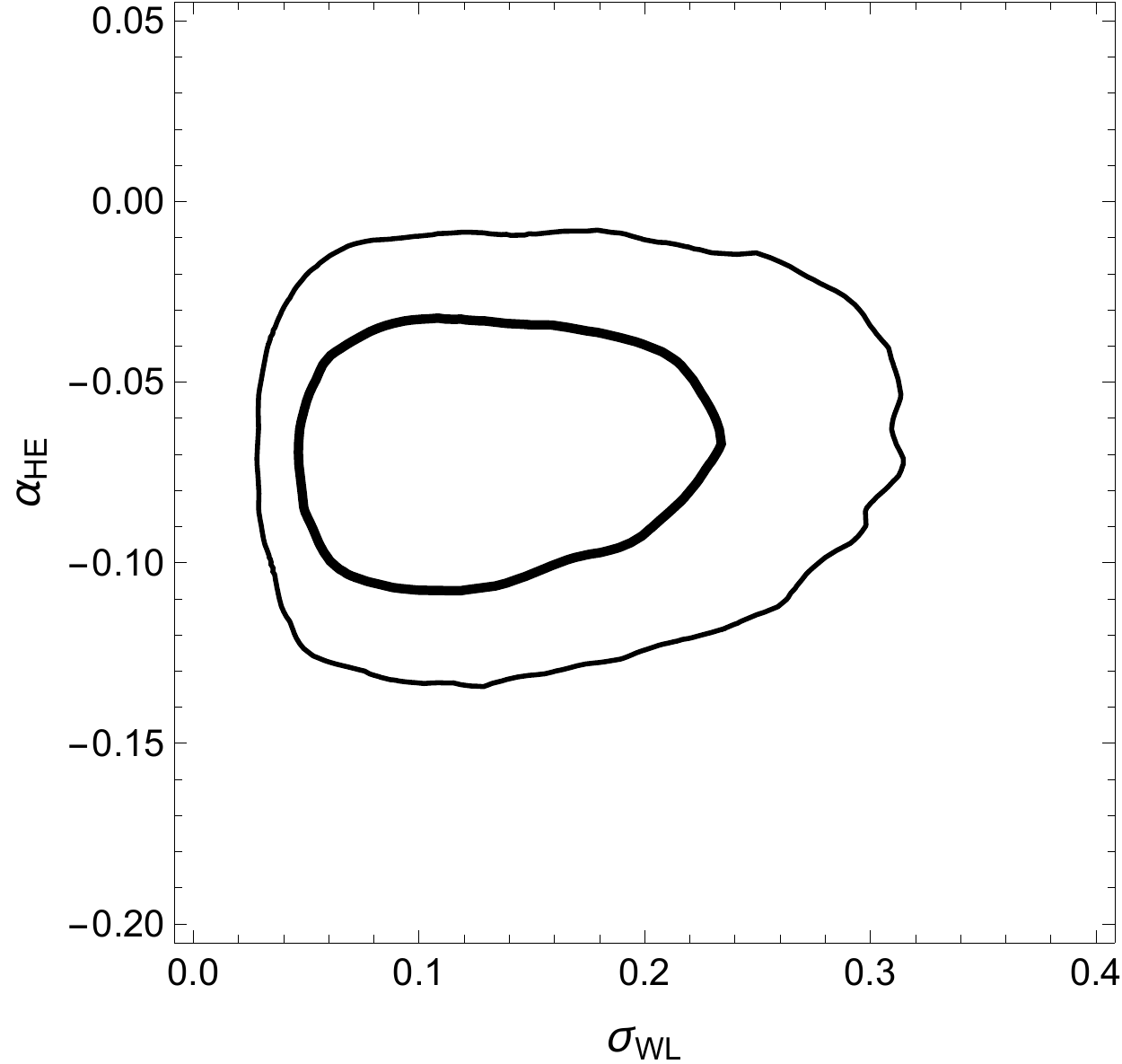} &
\includegraphics[width=5cm]{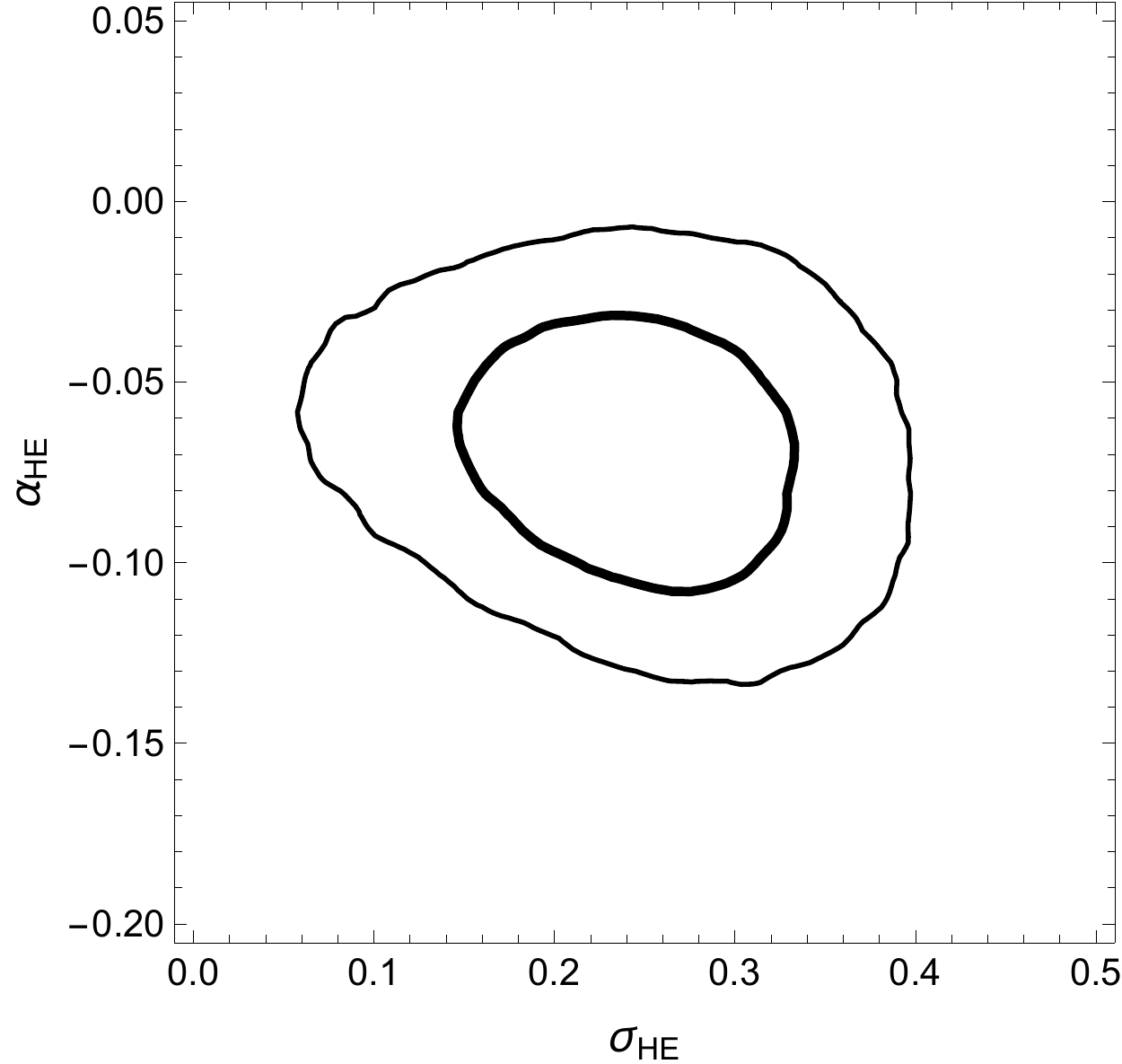} &
\includegraphics[width=4.6cm]{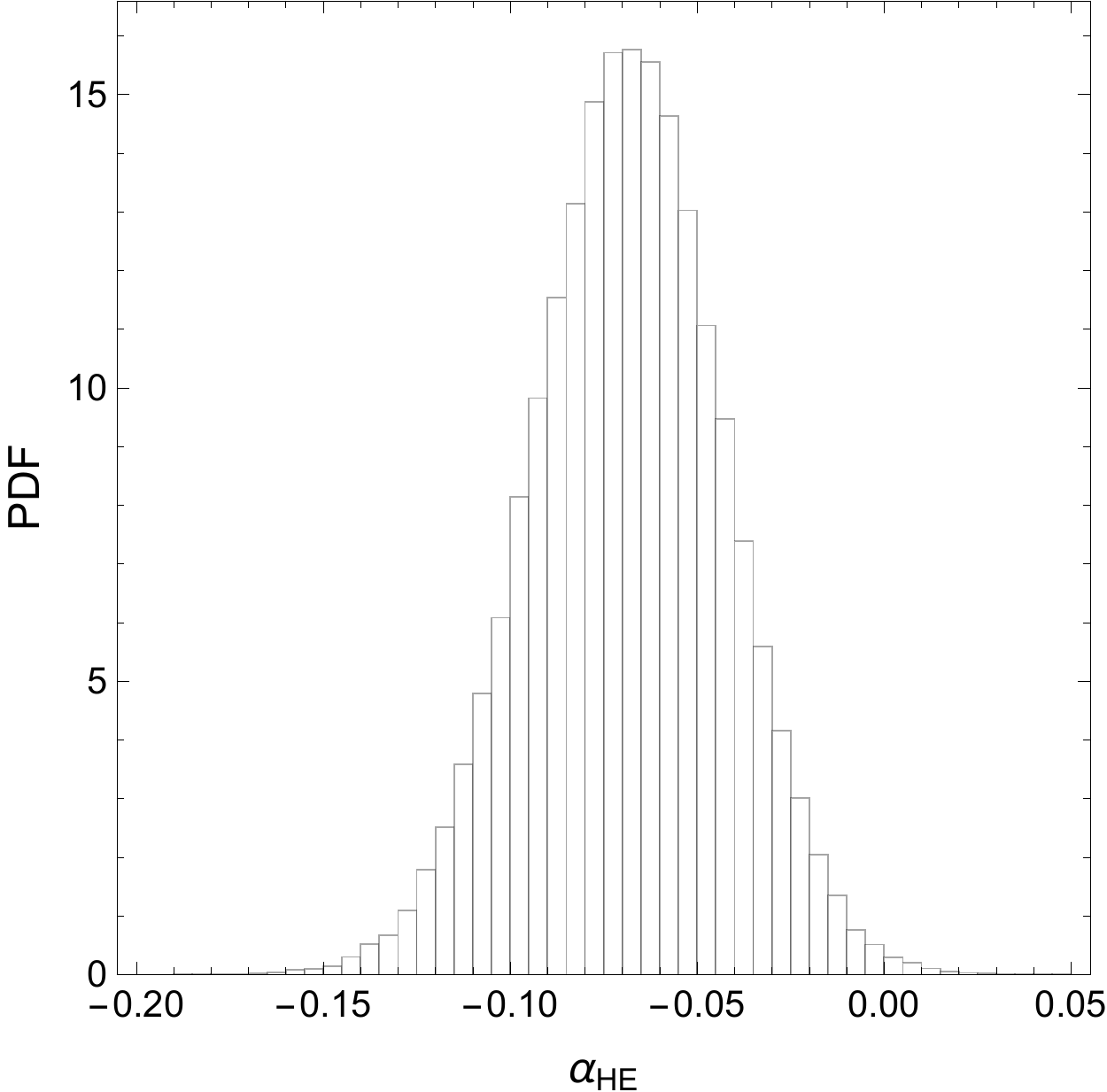}  \\
\includegraphics[width=4.8cm]{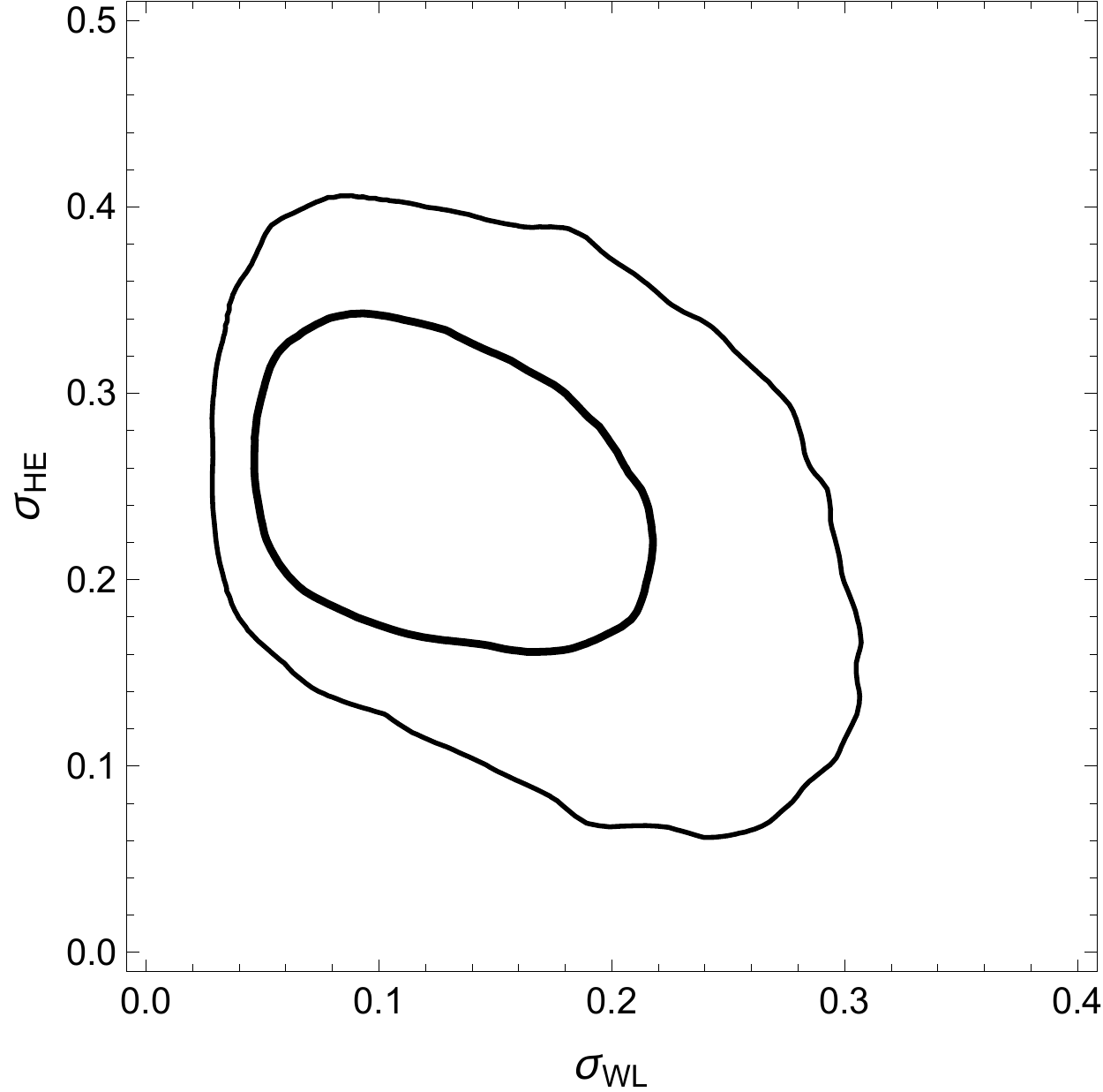} &
\includegraphics[width=4.6cm]{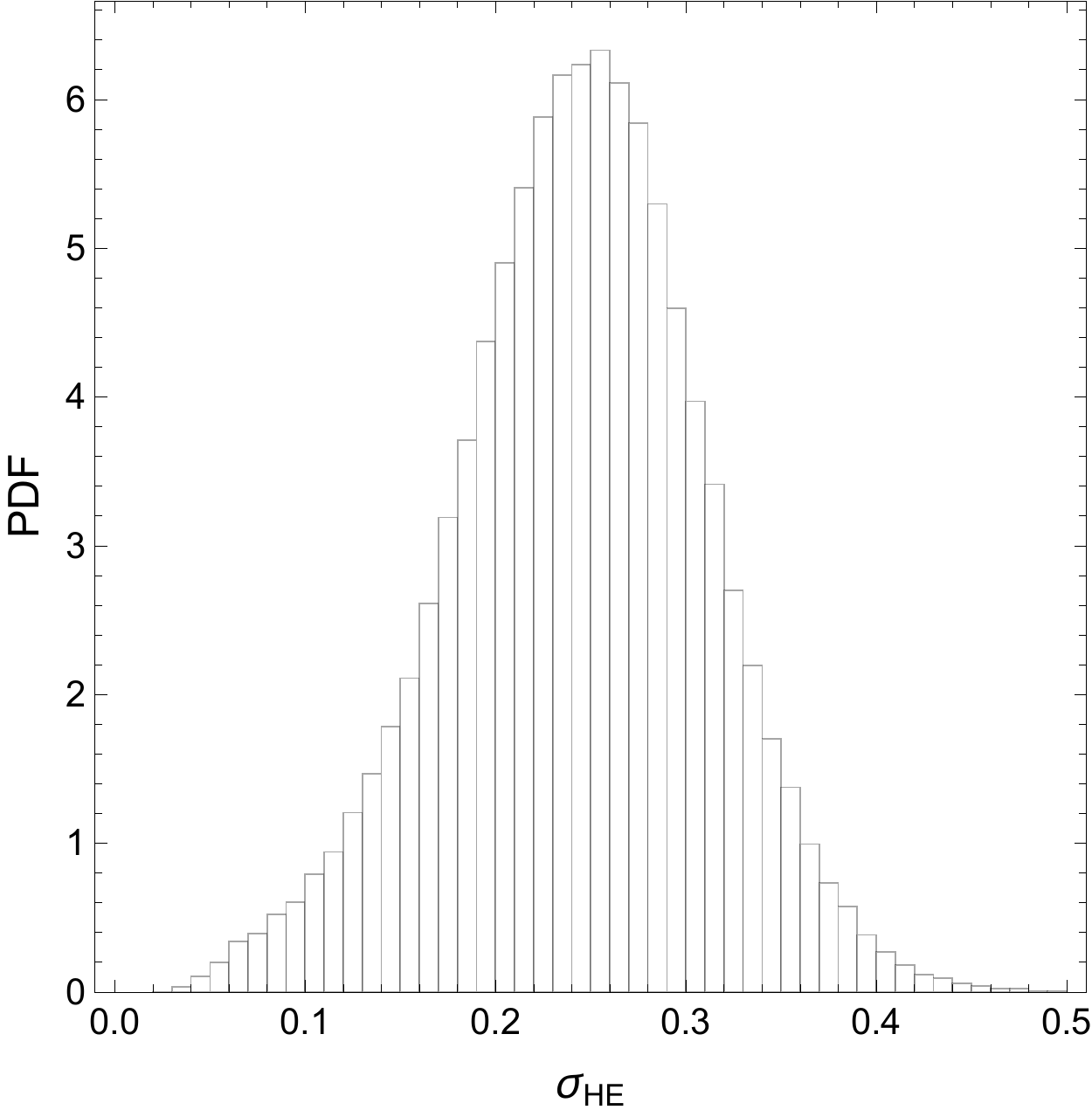} &  \\
\includegraphics[width=4.6cm]{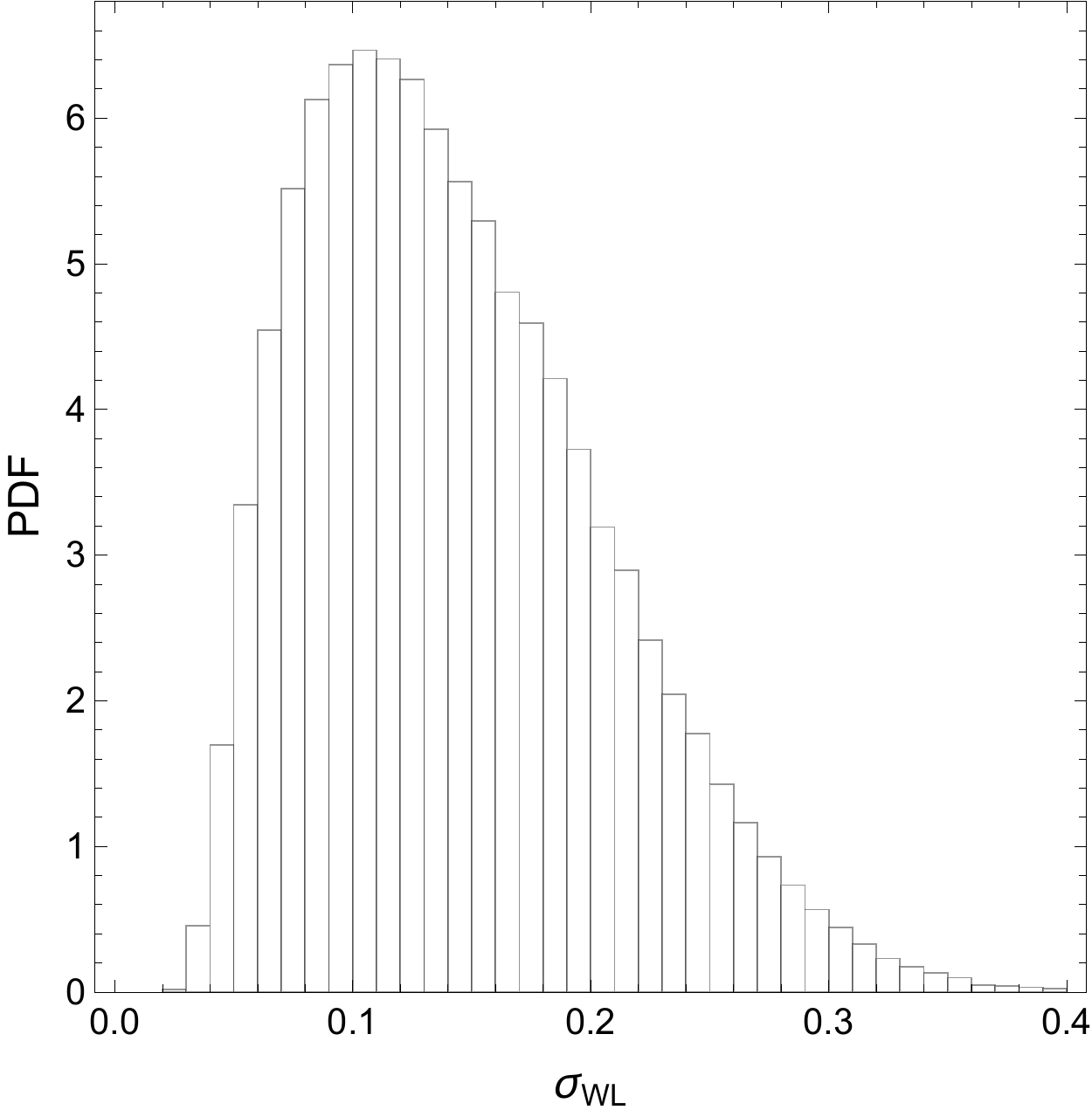} & & \\
\end{tabular}
\caption{Probability distributions of the scatter $\sigma_\mathrm{WL}$ of WL masses, of the scatter $\sigma_\mathrm{HE}$ of HE masses, and of the logarithm of the effective bias, $\alpha_{HE}=\ln M_{500}^\mathrm{HE}/M_{500}^\mathrm{WL}$, for the CCCP sample. HE masses were measured within $r_{500}^\mathrm{WL}$, the over-density radius related to the WL mass. The thick (thin) lines in the two-dimensional plots include the 1-(2-)$\sigma$ confidence region in two dimensions, here defined as the region within which the value of the probability is larger than $\exp(-2.3/2)$ ($\exp(-6.17/2)$) of the maximum.}
\label{fig_CCCP_PDF}
\end{figure*}

\begin{figure*}
\begin{tabular}{cc}
\includegraphics[width=8.5cm]{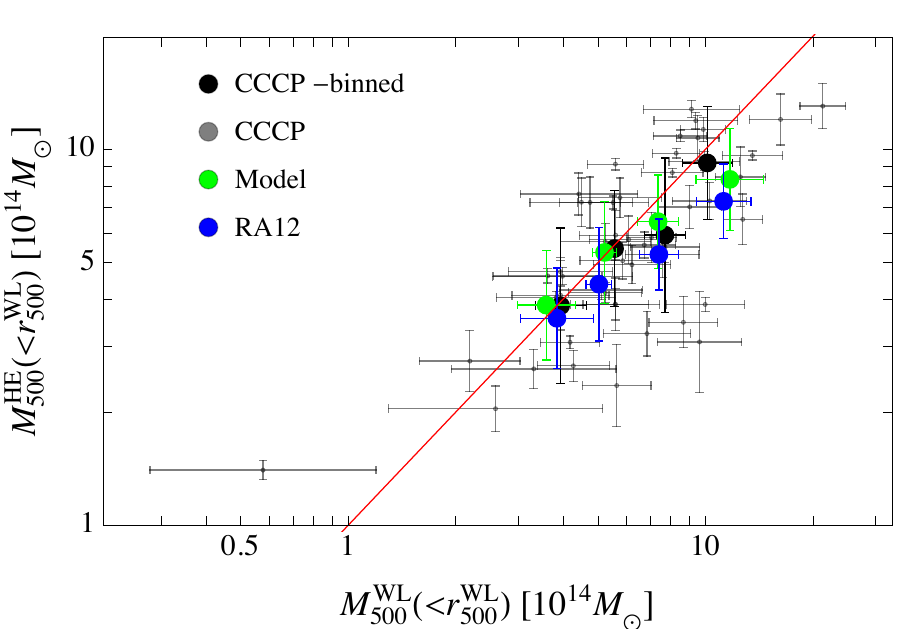} &\includegraphics[width=8.5cm]{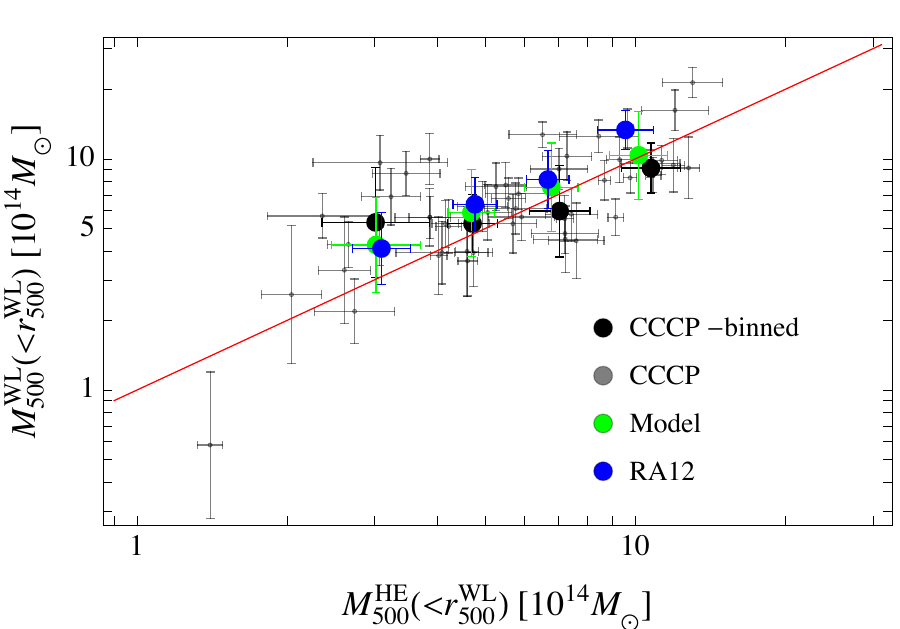}\\
\noalign{\smallskip}  
\includegraphics[width=8.5cm]{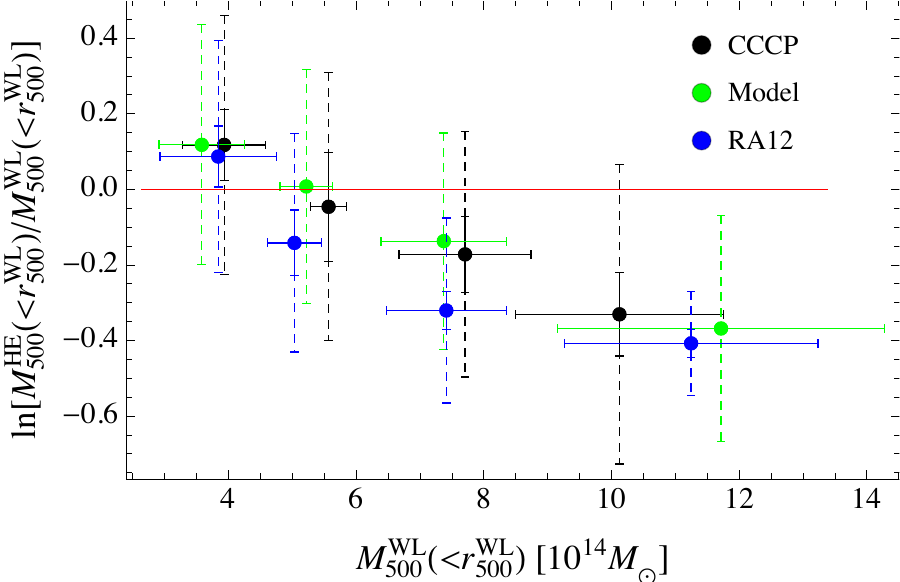} & \includegraphics[width=8.5cm]{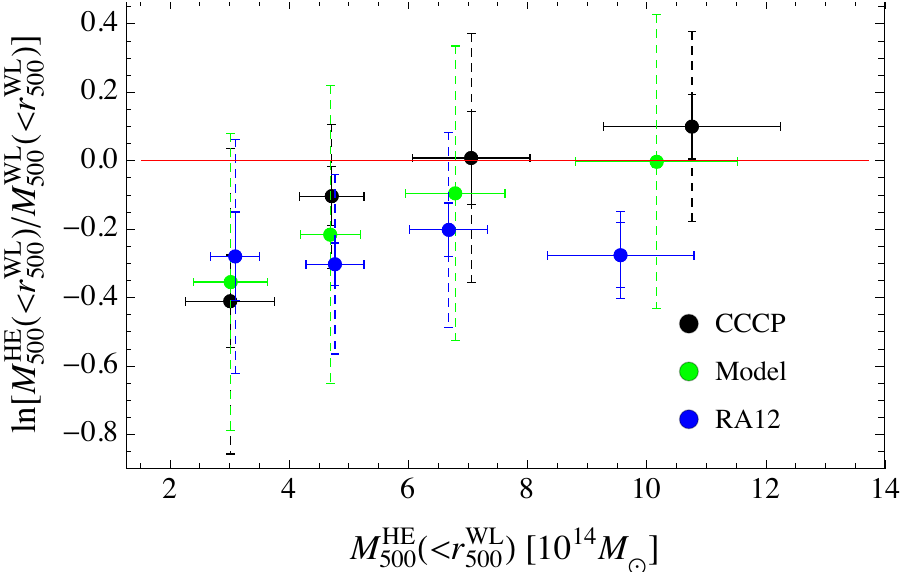}
\end{tabular}
\caption{Same as Fig.~\ref{fig_RA12_MWL_MHE} but for the clusters in the CCCP sample. Masses are measured within $r_{500}^\mathrm{WL}$, the over-density radius related to the WL mass. Blue points plot the results for clusters in the RA12 sample grouped in the same mass bins as the CCCP clusters. Masses for the RA12 sample are computed within $r_{500}^\mathrm{Tr}$.
}
\label{fig_CCCP_MWL_MHE}
\end{figure*}

\begin{figure*}
\begin{tabular}{cc}
\includegraphics[width=8.5cm]{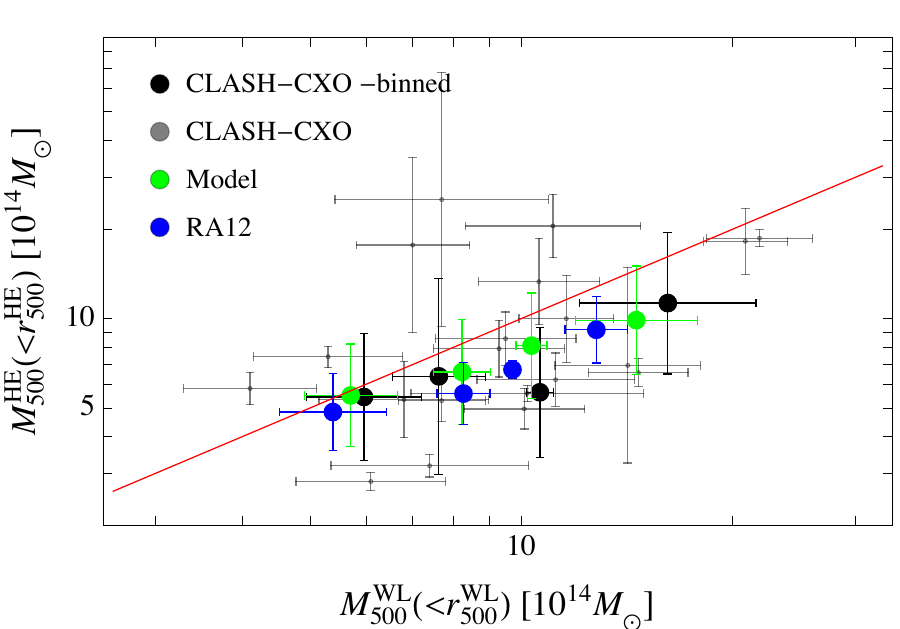} &\includegraphics[width=8.5cm]{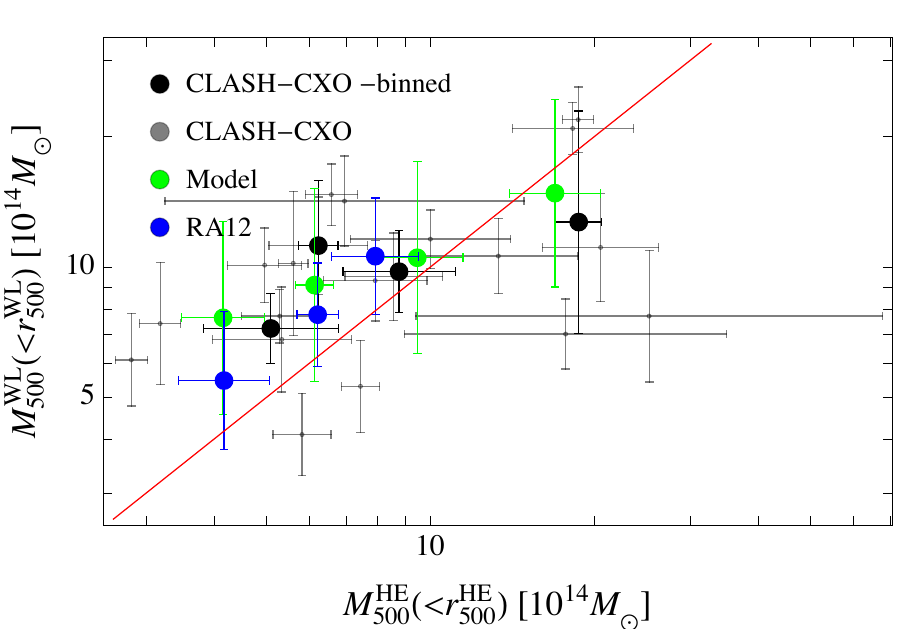}\\
\noalign{\smallskip}  
\includegraphics[width=8.5cm]{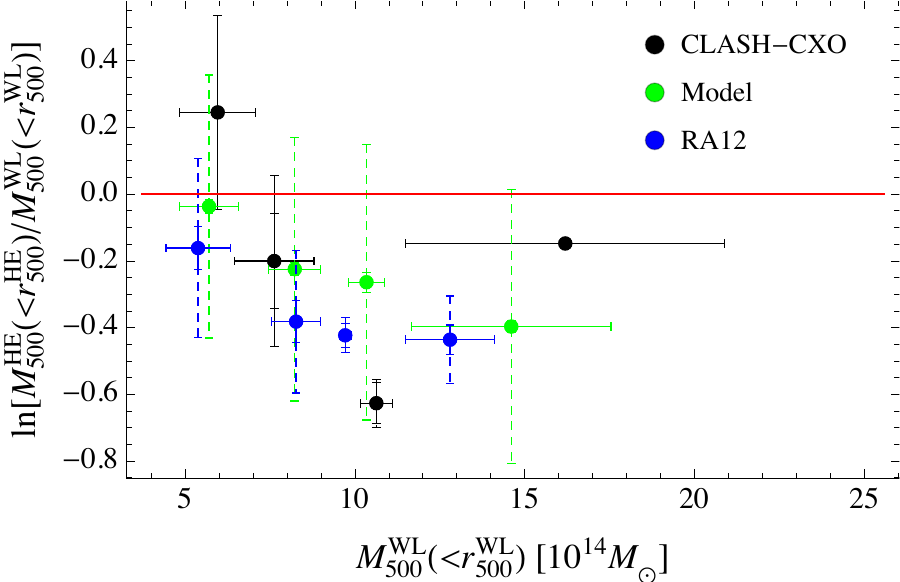} & \includegraphics[width=8.5cm]{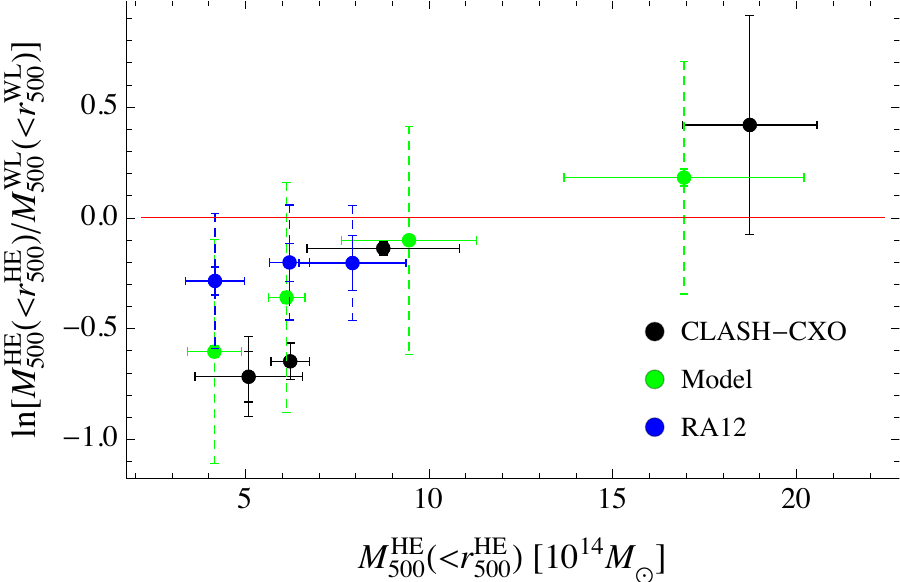}
\end{tabular}
\caption{Same as Fig.~\ref{fig_CCCP_MWL_MHE} but for the clusters in the CLASH sample with Chandra-based X-ray analyses. WL (HE) masses are measured within $r_{500}^\mathrm{WL}$ ($r_{500}^\mathrm{HE}$).
}
\label{fig_clashCHANDRA_MWL_MHE}
\end{figure*}

\begin{figure*}
\begin{tabular}{cc}
\includegraphics[width=8.5cm]{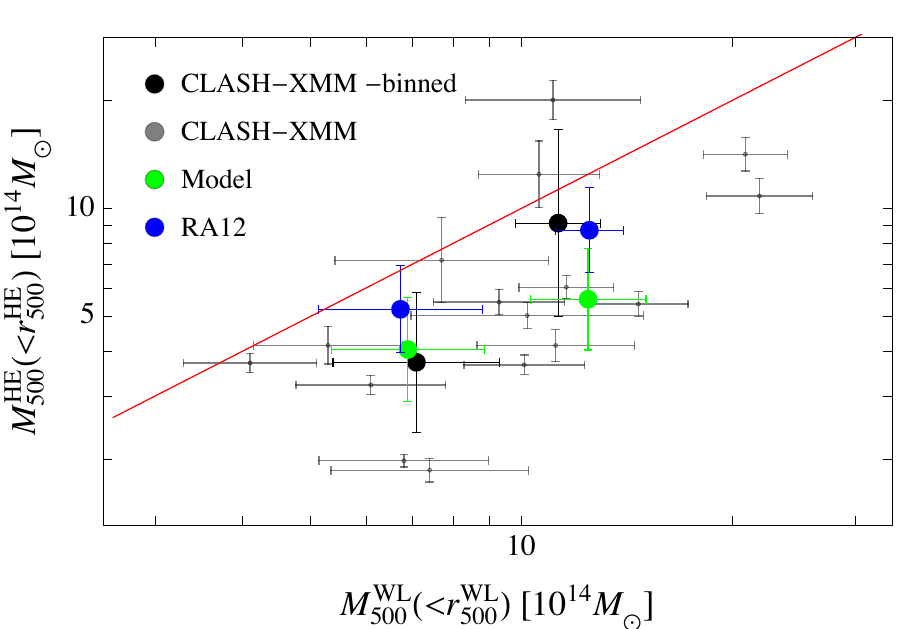} &\includegraphics[width=8.5cm]{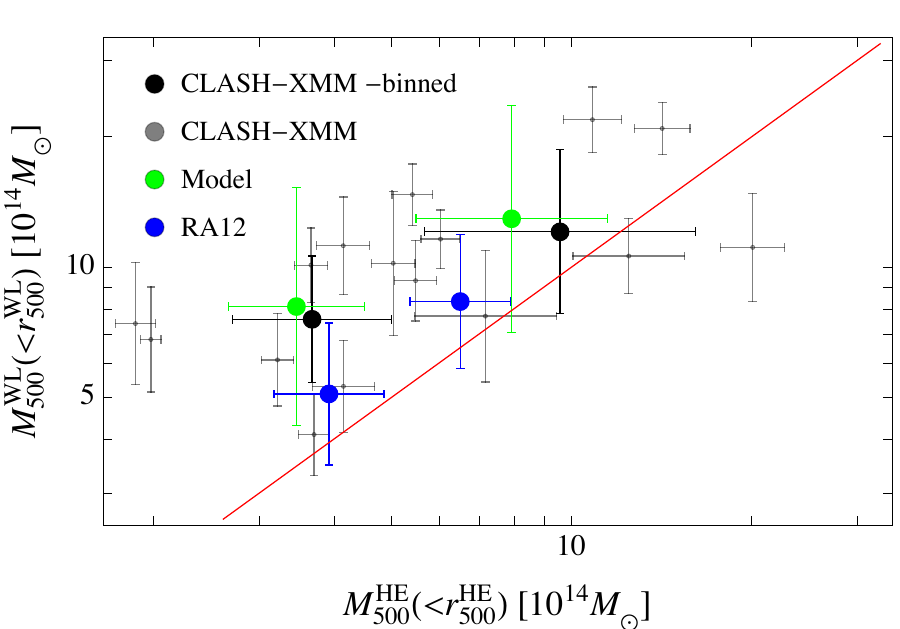}\\
\noalign{\smallskip}  
\includegraphics[width=8.5cm]{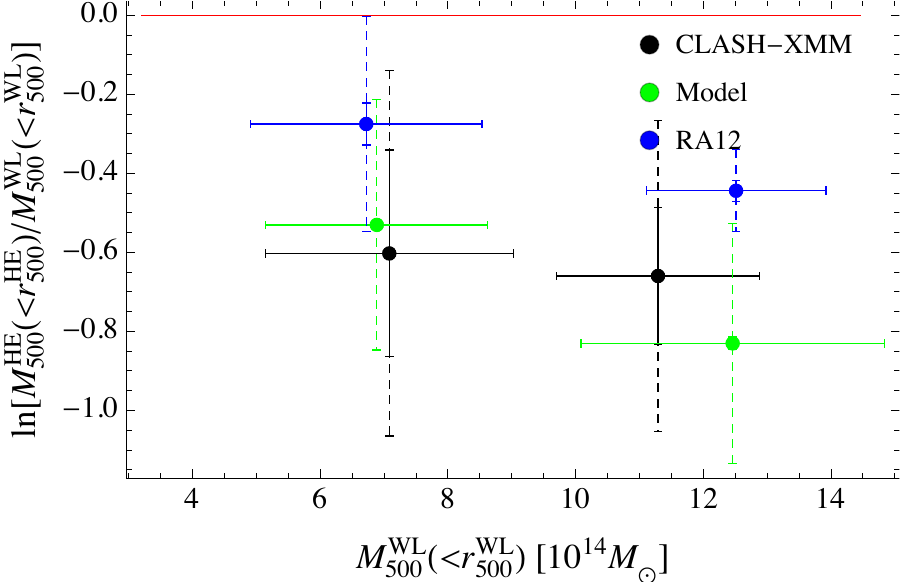} & \includegraphics[width=8.5cm]{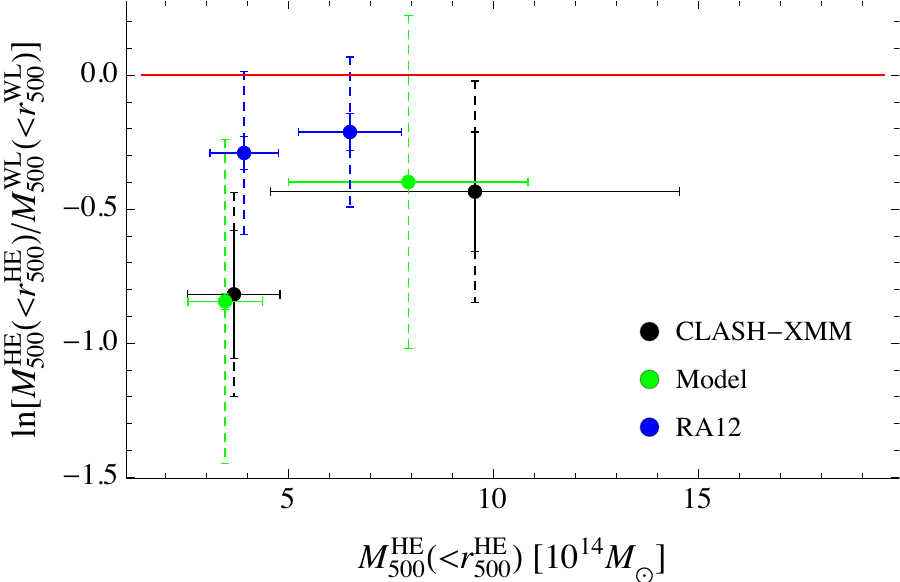}
\end{tabular}
\caption{Same as Fig.~\ref{fig_CCCP_MWL_MHE} but for the clusters in the CLASH sample with XMM-based X-ray analyses. WL (HE) masses are measured within $r_{500}^\mathrm{WL}$ ($r_{500}^\mathrm{HE}$).
}
\label{fig_clashXMM_MWL_MHE}
\end{figure*}

We considered a number of samples of clusters with observed WL and HE masses: $i$) the CCCP sample; $ii)$ the CLASH sample with X-ray estimates based on either Chandra or XMM data; $iii)$ the WTG clusters with HE masses from either B12 (WTG-B12) or L13 (WTG-L13). For the CCCP sample, we could consider either masses within the same radius, i.e., $r_{500}^\mathrm{WL}$, or alternatively WL masses within $r_{500}^\mathrm{WL}$ and HE masses within $r_{500}^\mathrm{HE}$. 

Results for the real clusters are summarised in Table~\ref{tab_scatter} and in Figs.~\ref{fig_CCCP_PDF},~\ref{fig_CCCP_MWL_MHE},~\ref{fig_clashCHANDRA_MWL_MHE}, and ~\ref{fig_clashXMM_MWL_MHE}. Thereafter in the text, when necessary we rescale the values of bias and scatters reported in Table~\ref{tab_scatter} to the corresponding values within the same length by simply multiplying by a factor 2/3. 

The parameter determination is not degenerate, see Fig.~\ref{fig_CCCP_PDF}. In fact, when comparing HE to WL masses, the relative intrinsic scatters acts in orthogonal directions.

HE masses are biased low with respect to WL masses by $\la 15$ per cent if we rely on the CCCP sample or by $\la$ 25--35 per cent if we consider the WTG estimates. The results for the CLASH sample depend on the X-ray analysis . The bias is $\sim$ 10 per cent for Chandra data and $\sim$ 30 per cent for XMM data.

The difference in the level of the bias among the various samples reflects the different absolute mass calibrations in the WL  and the X-ray samples, see Sec.~\ref{sec_mass_comp}. The bias ascertained with either the WTG or the CLASH-XMM sample is in agreement with results from numerical simulations whereas results based on the CCCP and CLASH-CXO slightly under-estimate it.

Apart from the overall normalisation, results from different data-sets are qualitatively and quantitatively consistent. The intrinsic scatter on WL masses is of order of $\sim$10-15 per cent, in very good agreement with numerical simulations. On the other hand, the estimated scatter on HE masses is $\sim$25 per cent, a factor of two larger than theoretical predictions. The large value of $\sigma_\mathrm{HE}$ is evident in the plots of $M_{500}^\mathrm{WL}/M_{500}^\mathrm{HE}$ versus the HE mass. The observed ratio increases much more steeply than the simulated ratio in RA12.

For the CCCP and the CLASH samples, we could restrict the analysis to either cool-core (CC) clusters or systems with low offsets between the BCG and the X-ray surface brightness peak. The HE mass of CC clusters in the CCCP sample is less biased. Due to the limited number of clusters, we could not confirm this result for the CLASH sample. There is no evident trend for the low off-sets clusters.

We verified a posteriori how well the statistical model reproduce the observed trends of the mass ratio, see Figs.~\ref{fig_CCCP_MWL_MHE},~\ref{fig_clashCHANDRA_MWL_MHE},~\ref{fig_clashXMM_MWL_MHE}. We generated a number of `true' masses from the normal distributions derived in the regression analyses and the corresponding `true' latent WL and HE masses, scattered according to the measured $\sigma_\mathrm{WL}$ and  $\sigma_\mathrm{HE}$. The observed manifest WL and HE masses were finally generated considering the measured statistical uncertainties. The masses were then binned as the real clusters. The observed trends in bias and scatter are well recovered. This test further validates a posteriori the assumptions of the Bayesian modelling, in particular that scatters are approximately log-normal as well as that the mass distribution can be described as a Gaussian.

The observed trends were also compared to the numerical simulations, see Figs.~\ref{fig_CCCP_MWL_MHE},~\ref{fig_clashCHANDRA_MWL_MHE}, and ~\ref{fig_clashXMM_MWL_MHE}. The extent of this comparison is limited. Firstly, the simulated clusters are not representative of the observed samples, even though both observed and simulated clusters should sample the tail at large values of the halo mass function. The discrepancy is further mitigated since we compared clusters in the same mass bins. Nevertheless, the dynamical and morphological properties of the simulated clusters may differ from the observed ones. Secondly, WL and HE masses of the simulated clusters were measured within the true $r_{500}$ differently from the real clusters whose over-density radii were estimated based on the WL/HE mass. This may limit the scatter of the simulated clusters compared to the real ones. 

Due to these differences only qualitative considerations can be made. Nevertheless some trends seem to further validate the picture detailed before. In particular, the observed mass bias $M^\mathrm{HE}/M^\mathrm{WL}$ scales with the WL mass as in simulations, which further supports that the scatter of measured WL masses agree with theoretical predictions. On the other hand, the observed mass bias as a function of the HE mass is significantly steeper than that in the simulated sample, which points to a larger than theoretically predicted scatter in HE masses.

\section{Discussion}
\label{sec_disc}

The main sources of bias and scatter in WL mass measurements are due to the presence of substructures and triaxiality \citep{ras+al12,gio+al14}. These effects are dominated by the dark matter component and are more easily reproduced in numerical simulations than the more complex processes involving gas physics. Reassuringly, the level of scatter we ascertained from observations is in very good agreement with the theoretical prediction of  $\sigma_\mathrm{WL}\ga 10$ per cent.

The observed relative bias between HE and WL masses is consistent within the statistical errors with predictions from simulations \citep{ras+al12}. Based on a suite of different numerical simulations \citep{bat+al12,kay+al12}, \citet{planck_2013_XX} estimated $b=1-M^{500}_\mathrm{HE}/M^{500}_\mathrm{Tr}=0.2^{+0.1}_{-0.2}$. However, the inherent uncertainty in the HE/WL calibrations prevents firmer conclusions. For some samples, the bias is as large as $\sim$50 per cent.

The measured intrinsic scatter in HE masses is larger than the theoretical prediction. The disagreement may hinge on several plausible causes. The formal statistical uncertainty in X-ray mass estimates is usually of the order of $\sim$10-15 per cent. However, the observed discrepancies among mass estimates from independent analyses are as large as 45 per cent ($\sim 30$ per cent within the same physical radius). The under-estimation of the formal error on the HE masses could determine an over-estimation of the intrinsic scatter.

\subsection{Gas physics}

Scatter in simulations may be under-estimated due to their current limits \citep{ras+al14}. Estimates of bias and scatter from numerical simulations are still uncertain, showing dependences on the physical treatment of the gas, and, possibly, on the hydrodynamical scheme adopted. Each simulation suite has circumstantial prescriptions for gas physics. Different treatments of radiative cooling and cooling/heating from the UV background play an important role. Thermal conduction in hot clusters may be effective in removing cold blobs and in making the thermal structure of the ICM more homogeneous. This leads to an increase of the spectroscopic temperature and therefore of the hydrostatic mass. Feedback from active galactic nuclei and supernovae can significantly reduce the temperature inhomogeneity. 

The impact of each ingredient is significant and each process may be more or less effective in different clusters. Theoretical predictions based on specific descriptions may then significantly under-estimate the intrinsic scatter in the HE mass.

\subsection{Simulation scheme}

Some disagreement among theoretical predictions is also caused by the adopted simulation scheme. Smoothed-particle-hydrodynamics (SPH) simulations produce larger temperature variations connected to the persistence of both substructures and their stripped cold gas than adaptive-mesh-refinement (AMR) codes \citep{sij+al11,ras+al14}, which lead to a more efficient mixing of gas entropy.  Low-entropy gas residing in high-density clumps is more efficiently mixed to the high-entropy ICM than in SPH simulations. The simulated temperature distribution is then more homogenous and the relative bias introduced in the estimate of X-ray temperature is smaller \citep{vaz+al11}. Around $r_{500}$, the temperature inhomogeneities of the SPH simulations can generate twice the typical hydrostatic-equilibrium mass bias of the AMR sample \citep{ras+al14}. 

These variations between simulation schemes make predictions less certain. A better understanding of the physical processes responsible for the complex thermal structure in ICM requires improved resolution and high sensitivity observations, first of all for higher temperature systems and larger cluster-centric radii \citep{ras+al14}.

\subsection{Mass dependence}

A further source of disagreement might be ascribed to any dependence of the bias on cluster mass. Neglecting such dependence can inflate the estimate of the scatter. The massive objects are expected to be the most disturbed ones, and they should have a complex temperature structure \citep{ras+al12}. This would imply a bias larger for the more massive clusters. We tested this hypothesis by repeating the analysis of Sec.~\ref{sec_resu} without fixing the slope $\beta_\mathrm{HE}$ to unity. 

Due to the addition of a new free parameter to be determined with the regression, we could obtain well constrained results only for the two richer samples. For the CCCP sample (all masses within $r_{500}^\mathrm{WL}$) , we obtained $\sigma_\mathrm{WL}=0.17\pm0.07$, $\beta_\mathrm{HE}=1.19\pm0.24$ and $\sigma_\mathrm{HE}=0.20\pm0.09$. For the CLASH-CXO sample, we obtained $\sigma_\mathrm{WL}=0.21\pm0.10$, $\beta_\mathrm{HE}=1.29\pm0.63$ and $\sigma_\mathrm{HE}=0.30\pm0.16$. 

For both samples, with respect to the results obtained fixing  $\beta_\mathrm{HE}=1$, the measured $\sigma_\mathrm{WL}$ is slightly larger whereas $\sigma_\mathrm{HE}$ is smaller. However, $\sigma_\mathrm{HE}$ is still larger than both $\sigma_\mathrm{WL}$ and the scatter predicted by numerical simulations. 

The estimated $\beta_\mathrm{HE}$ is slightly larger than unity, but still consistent within the statistical uncertainty. This scenario would then imply a still larger than expected scatter in HE masses at the expense of a not so plausible bias decreasing with mass ($\beta_\mathrm{HE}>1$). This alternative scenario is then more complex but does not solve the main incongruences it was supposed to address. Since the estimated $\beta_\mathrm{HE}$ is consistent with unity within the errors, we then disfavor this scenario.

\section{Conclusions}
\label{sec_conc}

In this paper, the first in a series which aims to critically revise the status quo in measuring cluster masses and calibrating scaling relations, we studied the biases and the intrinsic scatters of weak lensing and hydrostatic masses. Either WL or HE masses determined from different groups may differ by $\sim 40$ per cent, which hinders the absolute calibration of any scaling relation and the assessment of the relative bias between WL and HE masses.

We found that the intrinsic scatter of WL masses is of the order of $\sim$10-15 per cent, in line with theoretical predictions. The intrinsic scatter of HE masses turned out to be larger, $\sim$20-30 per cent, at odds with results from numerical simulations. The discrepancies may hinge on under-estimated statistical uncertainties in HE masses. A better understanding of the physical processes responsible for the complex thermal structure in the ICM and improved simulation schemes are also required to improve the theoretical predictions.

Most of the sources of scatter in the estimates of WL and HE masses are of well known origin. The assumption of spherical symmetry causes an over- or under- estimate of the WL mass whether the cluster is elongated in the plane of the sky or towards the observer, respectively. Departures from hydrostatic equilibrium or the difficult assessment of non-thermal contribution to the pressure limit the accuracy of HE masses. 

Over-simplified modelling inflates the intrinsic scatter. The joint analysis of multi-wavelength observations, from the X-ray to the optical band to the SZ effect in the radio, can provide unbiased estimates of the cluster mass \citep{def+al05,ser+al06,ser+al12a,ser+al13,mor+al12,lim+al13}. In fact, the combined information from the different data-sets enables us to recover the triaxial structure and the orientation of the cluster and to quantify the non-thermal contributions to the pressure.

An alternative approach is focusing on well-behaved clusters where bias and scatter are intrinsically small. Biases are lower in morphologically regular and isolated clusters \citep{ras+al12}. However, there are a few of them and they are rare to find. Even apparently spherical clusters with a regular morphology might significantly deviate from hydrostatic thermal equilibrium \citep{ser+al13}. Furthermore, a projected circular shape is well suited to either spherical systems or strongly prolate haloes elongated along the line of sight towards the observer.

Scatter and bias in WL and X-ray estimates play a fundamental role in the calibration of mass proxies. Ongoing programs are making significant efforts to understand the sources of systematics and to solve the related calibration issues \citep{roz+al14c,don+al14}. We quantified the size of the intrinsic scatter in WL and HE masses and discussed the effect of scatter in the determination of scaling relations. The scatter makes relation systematically flatter and more scattered. Proper statistical treatments could and should account for this.

\section*{Acknowledgements}
The authors thank Lauro Moscardini and Elena Rasia for useful discussions. MS acknowledges financial contributions from contracts ASI/INAF n.I/023/12/0 `Attivit\`a relative alla fase B2/C per la missione Euclid', PRIN MIUR 2010-2011 `The dark Universe and the cosmic evolution of baryons: from current surveys to Euclid', and PRIN INAF 2012 `The Universe in the box: multiscale simulations of cosmic structure'. SE acknowledges the financial contribution from contracts ASI-INAF I/009/10/0 and PRIN-INAF 2012 `A unique dataset to address the most compelling open questions about X-Ray Galaxy Clusters'.


\setlength{\bibhang}{2.0em}

\appendix

\section{Bias}
\label{app_bias}
The intrinsic scatter biases the average of the intrinsic variables with respect to the observable proxy. Here, we focus on intrinsic scatter and we assume that observational uncertainties are negligible. Let us consider a proxy $X$ of the independent variable $Z$,
\beq
\label{eq_bias_0}
X = \alpha_{X|Z} +\beta_{X|Z} Z \pm \sigma_{X|Z},
\eeq
where $\alpha_{X|Z}$ is the bias, $\beta_{X|Z}$ accounts for rescaling, and $\sigma_{X|Z}$ is the intrinsic scatter of the linear relation. We assume that the scatter is Gaussian,
\beq
\label{eq_bias_1}
p(X|Z) \propto \exp \left[ -\frac{1}{2}\left(\frac{X-(\alpha_{X|Z} +\beta_{X|Z} Z)}{\sigma_{X|Z}} \right)^2 \right].
\eeq

The intrinsic scatter can bias average values if the true variables are not uniformly distributed. The average intrinsic value of an ensemble of objects with the same measured $X$ can differ from $X$. The average is given by
\beq
\label{eq_bias_2}
\langle  Z \rangle (X) \propto \int Z \, p(X,Z) d Z.
\eeq
The analytical treatment of the bias is particularly simple when the distribution of the intrinsic variable is Gaussian,
\beq
\label{eq_bias_3}
p(Z) \propto \exp \left[ -\frac{1}{2}\left(\frac{Z-\mu_Z}{\sigma_{Z}} \right)^2 \right].
\eeq
A simple application of the Bayes' theorem shows that $p(X)$ is normally distributed too, with the same mean $\mu_Z$ and standard deviation 
\beq
\sigma_X=\sqrt{\sigma_{Z}^2+\sigma_{X|Z}^2}.
\eeq
In this case, the integral in Eq.~(\ref{eq_bias_2}) can be solved in terms of simple functions,
\beq
\label{eq_bias_4}
\langle  Z \rangle (X)= b_{\sigma} X +\Delta b_{\sigma},
\eeq
where $b_{\sigma}$ is the multiplicative bias due to the intrinsic scatter, 
\beq
\label{eq_bias_5}
b_{\sigma} =\frac{1}{\beta_{X|Z}}\frac{1}{1+\sigma_{X|Z}^2/(\beta_{X|Z} \sigma_Z)^2}, 
\eeq
and $\Delta b_{\sigma}$ is an additive bias,
\beq
\label{eq_bias_5b}
\Delta b_{\sigma} =  \left( \mu_Z \frac{\sigma_{X|Z}^2}{\beta_{X|Z}^2 \sigma_Z^2}-\frac{\alpha_{X|Z}}{\beta_{X|Z}} \right) \frac{1}{1+\sigma_{X|Z}^2/(\beta_{X|Z} \sigma_Z)^2} .
\eeq
The contribution to the bias due to intrinsic scatter is negligible either if the intrinsic scatter is very small ($\sigma_{X|Z}\rightarrow 0$) or if the intrinsic variable is uniformly distributed ($\sigma_{Z}\rightarrow \infty$). 

If we have to calibrate a linear relation,
\beq 
\label{eq_bias_6}
Y =\alpha_{Y|Z} +\beta_{Y|Z} Z ,
\eeq
but we have only measurements of the proxy $X$ instead of $Z$, we can not just study the relation
\beq
\label{eq_bias_7}
Y =\alpha_{Y|X} +\beta_{Y|X}   X ,
\eeq
and take $\beta_{Y|X}  $ as un unbiased estimator of $\beta_{Y|Z}$. In the Gaussian case,
\begin{eqnarray}
\label{eq_bias_8}
\alpha_{Y|X} & = & \alpha_{Y|Z} +  \Delta b_{\sigma} \beta_{Y|Z} , \\
\beta_{Y|X}  & = &  b_{\sigma} \beta_{Y|Z}  .
\end{eqnarray}
The intrinsic scatter apparently flattens the slope of the relation. The most convenient statistical approach requires a proper treatment of the selection effects and of the intrinsic scatter in the linear regression. De-biasing the data as suggested in Eq.~(\ref{eq_bias_4}) would correct each measured scattered proxy by the expected mean bias instead of the actual one, which is random for objects with the same measured $X$. Nevertheless this mean correction may provide a useful tool to quickly evaluate the effect of the intrinsic scatter. 

If the intrinsic scatter is log-normal, as it is usually the case, in the above discussion $X$ can be read, for example, as $\ln M^\mathrm{WL}$ or $\ln M^\mathrm{HE}$, whereas $Z$ stands for $\ln M^\mathrm{Tr}$.

\section{BCES}
\label{app_bces}

The Bivariate Correlated Errors and Intrinsic Scatter (BCES) method is a well known regression technique with good performance for data-sets with heteroscedastic and correlated errors on both axes as well as intrinsic scatter in the linear relation \citep{ak+be96}. The slope of the conditional linear relation in Eq.~(\ref{eq_bias_7}) can be estimated as
\beq
\label{eq_bces_1}
\beta(Y|X)=\frac{\sum_i (y_i-\langle y \rangle)(x_i-\langle x \rangle)-\sum_i \delta_{xy,i}}{
\sum_i (x_i-\langle x \rangle)^2-\sum_i \delta_{x,i}^2
},
\eeq
where $x_i$ ($y_i$) is the observed values of $X_i$ ($Y_i$), with associated observational uncertainty $\delta_{x,i}$ ($\delta_{y,i}$), and $\delta_{xy,i}$ is the covariance between errors; $\langle x\rangle$ and $\langle y\rangle$ are the mean values.

If the variable $X$ is scattered too, see Eq.~(\ref{eq_bias_0}), the slope in Eq.~(\ref{eq_bces_1}) is a biased estimator of $\beta_{Y|Z}$. For $\alpha_{X|Z}=0$ and $\beta_{X|Z}=1$, the unbiased slope is
\beq
\label{eq_bces_2}
\beta(Y|Z)=\frac{\sum_i (y_i-\langle y \rangle)(x_{i}-\langle x\rangle)-\sum_i \delta_{xy,i}}{
\sum_i (x_i-\langle x\rangle)^2-\sum_i (\delta_{x_i}^2 +\sigma_{X|Z}^2)
}.
\eeq
The modified BCES estimator should be used to evaluate the slope of the conditional linear relation if the covariate $X$ is scattered, i.e, $X$ is the scattered response variable of the covariate $Z$.

\section{Masses and cosmology}
\label{app_mass}

Masse estimates depend on the cosmological model. A conversion from other cosmological parameters may be required to convert to a reference model. The mass within a given cosmological over-density $\Delta$ is defined as
\beq
\label{eq_over_1}
M_\Delta =\frac{4\pi}{3}\Delta \rho_\mathrm{cr}(D_\mathrm{d} \theta_\Delta)^3,
\eeq
where $\theta_\Delta$ is the angular radius enclosing the overdensity and $D_\mathrm{d}$ is the angular diameter distance to the cluster.
 
Lensing 3D masses within a radius $r=D_\mathrm{d}\theta$, where $\theta$ is the aperture radius, scale as
\beq
\label{eq_over_2}
M^\mathrm{WL} \propto \Sigma_\mathrm{cr} (D_\mathrm{d} \theta_\mathrm{E}) D_\mathrm{d} ~\theta f(\theta)
\eeq
where  $\Sigma_\mathrm{cr}\equiv(c^2\,D_\mathrm{s})/(4\pi G\,D_\mathrm{d}\,D_\mathrm{ds})$  is the critical surface density for lensing, $ \theta_\mathrm{E}$ is the angular Einstein radius and $D_\mathrm{s}$ and $D_\mathrm{ds}$ are the source and the lens-source angular diameter distances, respectively. The function $f(\theta)\sim\theta^{\delta\gamma}$ quantifies the deviation of the mass profile from the isothermal case. At $r_{500}$, mass profiles are nearly isothermal, i.e., $\delta\gamma\sim 0$.

Solving for Eqs.~(\ref{eq_over_1}) and~(\ref{eq_over_2}), we obtain 
\begin{eqnarray}
M^\mathrm{WL}_\Delta  & \propto & D_\mathrm{d}^{-\frac{3\delta\gamma}{2-\delta\gamma}}\left(   \frac{D_\mathrm{ds}}{D_\mathrm{s}} \right)^{-\frac{3}{2-\delta\gamma}} H(z)^{-\frac{1+\delta\gamma}{1-\delta\gamma/2}} \\
& = & \left(  \frac{D_\mathrm{ds}}{D_\mathrm{s}} \right)^{-3/2}H(z)^{-1}~\mathrm{for~\delta\gamma=0}.
\end{eqnarray}

Hydrostatic masses within $\theta$ scales as
\beq
\label{eq_over_3}
M^\mathrm{HE} \propto D_\mathrm{d}~\theta^{1+\delta\gamma};
\eeq
the HE mass within a given cosmological over-density is then
\begin{eqnarray}
\label{eq_over_4}
M^\mathrm{HE}_\Delta & \propto &  D_\mathrm{d}^{-\frac{3\delta\gamma}{2-\delta\gamma}}H(z)^{-\frac{1+\delta\gamma}{1-\delta\gamma/2}} \\
& = & H(z)^{-1}~\mathrm{for~\delta\gamma=0}.
\end{eqnarray}
For $\delta\gamma=0$, $M^\mathrm{HE}_\Delta/r_\Delta$ is independent of the adopted cosmology. When it was required, we used the above relations with $\delta\gamma=0$ to make the proper conversion from different cosmological parameters. We refer to \citetalias{ser14_comalit_III} for further details. An analog treatment for X-ray observables can be found in \citet{man+al10}.

\end{document}